\documentclass[11pt, draftclsnofoot, onecolumn]{IEEEtran}
\usepackage[utf8]{inputenc}
\usepackage{cite}

\usepackage[cmex10]{amsmath}
\usepackage{amssymb,latexsym}
\usepackage{epsfig}
\usepackage{caption}
\usepackage{color}
\usepackage{graphicx}
\usepackage{subcaption}
\usepackage{bbm}% to get the indicator function to work
\usepackage{pst-plot}
\usepackage{pstricks-add}
\usepackage{breqn}
\usepackage{algpseudocode}
\usepackage{algorithm}
\usepackage{psfrag}
\usepackage{mathtools} %for bsmallmatrix

 \allowdisplaybreaks
\def\myrootdir{/Users/ayca/Dropbox/DRAFTS}

\def\bibdirM{\myrootdir/MY_BIB}
\def\bibdirMM{\myrootdir/MY_BIB}
\def\bibdirC{\myrootdir/MY_BIB}

\algnewcommand{\Initialize}[1]{%
  \State \textbf{Initialize:}
  \Statex \hspace*{\algorithmicindent}\parbox[t]{.8\linewidth}{\raggedright #1}
}

\newcommand{\err}{\ensuremath{\varepsilon}} % \myerr

\newcommand{\mysnr}{\ensuremath{\gamma}} % N_{Q_e} %\alpha_{snr}
\newcommand{\lw}{\ensuremath{l_w}} % N_{Q_e} %\alpha_{snr}

\newcommand{\newoperator}[3]{\newcommand*{#1}{\mathop{#2}#3}}

\newoperator{\tr}{\mathrm{tr}}{\nolimits}
\newoperator{\diag}{\mathrm{diag}}{\nolimits}
\newoperator{\rank}{\mathrm{rank}}{\nolimits}
\newoperator{\myperm}{\mathrm{perm}}{\nolimits}
\newcommand{\expectation}{\ensuremath{\mathbb{{E}}}} %expectation operator E

\newcommand{\tminus}{\ensuremath{{t_{-}}}}
\newcommand{\tplus}{\ensuremath{{t_{+}}}}

\newoperator{\myvec}{\mathrm{vec}}{\nolimits}

\newcommand{\herm}{\mathrm{\dagger}}%{\dagger}{H}

 \newcommand{\myQED}{$\square$}

\newtheorem{cor}{\bf{Corollary}}[section]
\newtheorem{lem}{\bf{Lemma}}[section]

\newtheorem{defn}{\bf{Definition}}[section]
\newtheorem{rem}{\bf{Remark}}[section]

\newenvironment{lemma}
{ \vspace{3pt} \par\noindent  \lem \begin{itshape}\noindent}
{\end{itshape} \vspace{6pt}}
  \newenvironment{definition}
{\vspace{3pt} \par\noindent \defn \begin{itshape}\noindent}
{\end{itshape} \vspace{6pt}}
\newenvironment{remark}
{\vspace{2pt} \par\noindent \rem \begin{itshape}\noindent}
{\end{itshape}\vspace{2pt}}
\newenvironment{corollary}
{\vspace{3pt} \par\noindent \cor \begin{itshape}\noindent}
{\end{itshape}\vspace{6pt}}

\begin{document}

\title{Transmission Strategies for  Remote Estimation with an Energy Harvesting Sensor}

 \author{Ay\c ca \"Oz\c celikkale, Tomas McKelvey, Mats Viberg% <-this % stops a space
 \thanks{A.~\"Oz\c celikkale, T.~McKelvey and M.~Viberg are with Dep. of Signals and Systems, Chalmers University of Technology, Gothenburg, Sweden  e-mails: \{ayca.ozcelikkale, tomas.mckelvey, mats.viberg\}@chalmers.se. }
% %\thanks{}
% \\
 }
\date{}
\maketitle

\begin{abstract}
We consider the remote estimation of a time-correlated signal using an energy harvesting (EH) sensor.  The sensor observes the unknown signal and communicates its observations to a remote fusion center using an amplify-and-forward strategy. We consider the design of optimal power allocation strategies in order to minimize the mean-square error at the fusion center. Contrary to the traditional approaches, the degree of correlation between the signal values constitutes an important aspect of our formulation. We provide the optimal power allocation strategies for a number of illustrative scenarios.  We show that  the most majorized power allocation strategy, i.e. the power allocation as balanced as possible,  is optimal for the cases of circularly wide-sense stationary (c.w.s.s.) signals  with a static correlation coefficient, and  sampled low-pass c.w.s.s. signals for a static channel. We show that the optimal strategy can be characterized as a water-filling type solution for  sampled low-pass c.w.s.s. signals for a fading channel. Motivated by the high-complexity of the numerical solution of the optimization problem,  we propose low-complexity policies for the general scenario. Numerical evaluations illustrate the close performance of these low-complexity policies to that of the optimal policies,  and demonstrate the effect of the EH constraints and the degree of freedom of the signal.
\end{abstract}

\section{Introduction}\label{sec:intro}
Energy harvesting solutions offer a promising framework for future wireless sensing systems. Instead of completely relying on a fixed battery or power from the grid, nodes with EH capabilities can collect energy from the environment, such as solar power or power from mechanical vibrations.
In addition  to  enabling  energy autonomous sensing systems,   EH capabilities also offer  prolonged network life-times  and enhanced mobility for the nodes in the network  \cite{gunduzStamatiouMichelusiZorzi_2014,ulukusYenerErkipSimeoneZorziGroverHuang_2015}.

One of the key issues in the design of EH systems is the intermittent nature of the energy supply.
In a traditional device,  the energy that can be used for communications has either a fixed known value for each transmission or there is a total energy constraint. In contrast, for  an EH node, the energy available for information transmission  depends on the energy used in previous transmissions and the energy that may be available in the future.
In such systems, the transmission strategies have to be re-designed in order to ensure reliable and efficient operation in the entire time frame of interest. For instance, at a given instant, an EH node may have to choose between increasing the energy used in the current transmission to increase reliability at that instant or saving the energy for  upcoming transmissions due to forecasted poor energy harvesting conditions in the future.

In that respect, the problem of reliable communications with EH nodes  have been studied   under a broad range of scenarios  \cite{gunduzStamatiouMichelusiZorzi_2014,ulukusYenerErkipSimeoneZorziGroverHuang_2015,ozelUlukus_2012,DongFarniaOzgur_2015,ozelTutuncuogluYangUlukusYener_2011,tutuncuogluYener_2012,YangOzelUlukus_2012,antepliUysalErkal_2011,yangUlukus_2012mac}. 
Capacity of  point-to-point Gaussian channels are considered in \cite{ozelUlukus_2012,DongFarniaOzgur_2015}. Total throughput maximization and transmission time completion problems are investigated in  \cite{ozelTutuncuogluYangUlukusYener_2011,tutuncuogluYener_2012}. Multi-user scenarios have been considered, including broadcast channels \cite{YangOzelUlukus_2012,antepliUysalErkal_2011} and multiple-access channels  \cite{yangUlukus_2012mac}.
An overview of these recent advances in EH communication systems is provided in \cite{gunduzStamatiouMichelusiZorzi_2014,ulukusYenerErkipSimeoneZorziGroverHuang_2015}.
In contrast to the these works, whose focus is on the reliable communication problem, here we adopt an alternative approach and focus on the  estimation aspect of the problem, i.e. recovery of the unknown signal measured by the sensors.

At the moment, the literature on the estimation aspect, in particular investigations on the effect of the possible statistical correlation between the unknown signal values, is quite limited.
% %
 Previously, the degree of correlation of the unknown signal has been shown to have a substantial impact on the optimum sensor communication strategies  without EH constraints \cite{BahceciKhandani_2008, ShiraziniaDeyCiuonzoRossi_2016,ayca_IEEESP2010, ayca_unitaryIT2014}. 
 In the case of EH systems, only a limited number of works address this issue.
Optimal transmission strategies for the estimation of independently identically distributed (i.i.d.) Gaussian sources follow from the findings of  \cite{ozelUlukus_2012,OrhanGunduzErkip_2015,zhaoChenZhang_2015}.
Majorization based arguments of \cite{ozelUlukus_2012} show that energy allocations that are as balanced as possible are optimal for i.i.d. sources.
Estimation of i.i.d. sources is considered under a source coding perspective, and an associated 2-D water-filling interpretation is provided in \cite{OrhanGunduzErkip_2015}. A  water-filling type characterization of optimal solutions for uncoded transmission are provided by \cite{zhaoChenZhang_2015}.
The parameter estimation problems considered in \cite{nourianDeyAhlen_2015,knornDeyAhlenQuevedo_2015} provide insights about the limiting case, where the unknown value is fully spatially correlated across sensors. In particular, a threshold based policy is shown to be optimal under a binary energy allocation strategy  \cite{nourianDeyAhlen_2015}. Extensions of this framework, where energy sharing between sensors are possible, is provided in \cite{knornDeyAhlenQuevedo_2015}.
Investigations in \cite{NayyarBasarTeneketzisVeeravalli_2013,NourianLeongDey_2014,calvofullanaMatamorosAntonHaro_2015} provide guidelines for Markov sources.
%
%Under a binary decision rule of transmitting or not transmitting,
A threshold based strategy is found to be optimal for Markov sources where the sensor transmits if the difference between the current source value and the most recently transmitted value exceeds the threshold  \cite{NayyarBasarTeneketzisVeeravalli_2013}.
Optimal power allocations  for a vector Gaussian Markov source  under an unreliable channel with packet erasures is considered in \cite{NourianLeongDey_2014}.
A characterization of the optimal power allocations for temporally correlated Markov sources  is provided  in terms of water-filling type solutions under a source-coding framework in \cite{calvofullanaMatamorosAntonHaro_2015}.
A distributed source coding framework for spatially correlated sources is considered in \cite{GangulaGunduzGesbert_2015,tapparelloSimeoneRossi_2014}.

Here we focus on the estimation of a time-correlated  Gaussian signal using an EH sensor.
The EH sensor observes the unknown signal and communicates its observations to the remote fusion center under energy harvesting constraints.  We consider an amplify-forward strategy motivated by the the high computational and the energy cost of source and channel coding operations;  and the fact that for estimation of a Gaussian source, uncoded transmission (analog forwarding) is optimal for additive white Gaussian (AWGN) channels under mean-square error \cite{GastparRimoldiVetterli_2003,gastpar_2008}; which is also extended to energy harvesting scheme for i.i.d. Gaussian signals in the asymptotic regime \cite{zhaoChenZhang_2015}.
We focus on the problem of optimal power allocation in order to minimize the mean-square error (MSE) over a finite-length horizon at the fusion
center.
Here we consider a general fading channel scenario whereas an investigation for the static channel case with limited proofs is provided in \cite{ayca_eusipco2016}.

We adopt the off-line optimization scheme, where the sensor knows the energy arrivals and the channel gains acausally.
Off-line optimization approaches have been investigated  for various scenarios, such as  point-to-point channels\cite{ozelTutuncuogluYangUlukusYener_2011,tutuncuogluYener_2012}, broadcast channels \cite{YangOzelUlukus_2012,antepliUysalErkal_2011} and multiple-access channels \cite{yangUlukus_2012mac} under rate based performance criterion   as well as for source coding  \cite{OrhanGunduzErkip_2015, calvofullanaMatamorosAntonHaro_2015,GangulaGunduzGesbert_2015} and remote estimation scenarios\cite{nourianDeyAhlen_2015}. 
From an energy harvesting perspective, these type of approaches are well-suited  for  scenarios where the energy arrivals can be accurately predicted, such as RF energy harvesting scenarios with dedicated power transfer scheduling as in \cite{GuoWandYang_2014,duFischioneXiao_2016}. 
Off-line optimization approaches also provide benchmarks to evaluate the fundamental performance limitations for energy harvesting systems and  structural guidelines  which facilitate possibly sub-optimal but efficient solutions for the general case. Examples for this include the online near-optimal scheme of \cite{WangLiu_2013} which uses the off-line directional water-filling solution of \cite{ozelTutuncuogluYangUlukusYener_2011} and the block transmission scheme of \cite{ayca_isit2016} which is motivated by the most-majorized power allocation of   \cite{ozelUlukus_2012} optimal for the off-line scheme.

We provide the optimal power allocation strategies for a number of illustrative scenarios.
We present  water-filling type characterizations of the optimal strategies  for  uncorrelated sources. These characterizations make use of a time-index dependent threshold, which is a typical property of the EH solutions  \cite{ozelTutuncuogluYangUlukusYener_2011}.
For the parameter estimation case, i.e. fully correlated signal scenario, the strategy that only sends the data in the time slots with the most favorable channel conditions is shown to be optimal.
We also consider circularly wide-sense stationary signals, which constitute a finite dimensional analog of wide-sense stationary signals \cite{neeser_proper_1993,GrayToeplitzReview}.    
We note that, in general, the components of c.w.s.s. signals are possibly correlated and the calculation of mean-square error requires  a matrix inversion as opposed to a direct sum of rate functions as in the case of throughput based formulations \cite{tutuncuogluYener_2012,YangOzelUlukus_2012,antepliUysalErkal_2011}. 
Nevertheless, we show that water-filling type characterizations of optimal strategies also hold for sampled low-pass c.w.s.s. signals for fading channels.    
We also show that the most majorized power allocation strategy, i.e. the power allocation as balanced as possible,  is optimal regardless of the degree of correlation in the cases of c.w.s.s. signals with static correlation coefficient and  sampled low-pass c.w.s.s. signals for a static channel.  
Although one may expect that as the signal components become more correlated,  strategies that send a low number of signal components with higher power become optimal instead of strategies that allocate power as uniform as possible, the case of static correlation shows that this may not be always the case.

These results on c.w.s.s. signals complement the other  scenarios where  balanced power allocations are found to be optimal, in particular, the  i.i.d. sources scenario that follows from the findings of \cite{ozelUlukus_2012}  and  sensing of two correlated Gaussian variables studied in a rate-distortion framework in \cite{GangulaGunduzGesbert_2015}. 
We note that,  by definition, the covariance matrices associated with c.w.s.s. signals are circulant \cite{neeser_proper_1993,GrayToeplitzReview}. %
Due to the asymptotic equivalence of sequences of circulant and Toeplitz matrices, (which constitute the covariance matrices of wide-sense stationary signals \cite{GrayToeplitzReview}), our  investigations here can be considered as an intermediate step towards understanding limitations imposed by energy harvesting to sensing of wide-sense stationary signals, which is a fundamental signal model  in the fields of communications and signal processing.

 Motivated by  the high complexity of the numerical solution of the optimization problem for the general scenario, we propose a number of low-complexity policies.
  % Based on lower and upper bounds on the mean-square error,
 These policies are based on lower and upper bounds on the mean-square error and provide  possibly sub-optimal but nevertheless efficient approaches to the power allocation problem at hand. Numerical evaluations illustrate the close performance of these low-complexity policies to that of the optimal policies,  and demonstrate the effect of the energy harvesting constraints and the degree of freedom of the signal on the system performance.

%%%%%%%%%%%%%%%%%%%%%%%%%%%%%%%%55
The rest of the paper is organized as follows. We present the problem formulation  in Section~\ref{sec:sys}.
In Section~\ref{sec:optpolicy}, the optimal strategies for a number of scenarios are provided.  In Section~\ref{sec:lowcomp},  low-complexity strategies for the general case are proposed.  Numerical evaluations are provided in Section~\ref{sec:num}. The paper is concluded in Section~\ref{sec:conc}.

%%%%%%%%%%%%%%%%%%%%%%%%%%%%%%%%%%%%%%%%%%%
{\it Notation:}
    The complex conjugate transpose of a matrix $A$  is denoted by $A^\herm$.  The i$^{th}$ row, k$^{th}$ column element of a matrix $A$ is denoted by $[A]_{ik}$. The positive semi-definite (p.s.d.) ordering for Hermitian matrices  is denoted by $\succeq$. $I_n $ denotes the identity matrix with $I_n \in \mathbb{C}^{n \times n}$.
% Uppercase and lowercase letters denote matrices, and column/row vectors, respectively.
%%%%%%%%%%%%%%%%%%%%%%%%%%%%%%%%%%%%%%%%%%%%%%%%%%%%%%%%

%\kern-0.2em
 \section{System Model and Problem Statement} \label{sec:sys}
%\kern-0.1em

\subsection{Signal Model}\label{sec:sys:sig}

The aim of the remote estimation system is to estimate the unknown complex proper zero-mean Gaussian signal ${\bf{x}}$ defined over time as $ {\bf{x}}=[x_1,\ldots, x_t,\ldots, x_{n}] \in  \mathbb{C}^{n \times 1}$, $ {\bf{x}} \sim \mathcal{CN}(0,{K_{\mathbf x}})$ with $K_{\mathbf x} =\expectation[\bf{x}  \bf{x}^\herm ]$, $P_x \doteq \tr[K_{\bf{x}}]$.
%Here we focus on the case where the signal values are not necessarily uncorrelated, hence the covariance matrix $K_{\mathbf{x}}$ is not necessarily identity.
We denote  the eigenvalue decomposition (EVD) of $K_{\bf{x}}$ as $K_{\bf{x}} = U \Lambda_{x}  U^\dagger$, where $\Lambda_{x} \in  \mathbb{R}^{n \times n}$ is the diagonal matrix of eigenvalues and $U \in \mathbb{C}^{n \times n}$  is a unitary matrix.
 Let $s$  with $s \leq n$ be the number of non-zero eigenvalues of $K_{\bf{x}}$, i.e. rank of $K_{\bf{x}}$. Let $\Omega$ denote the index set of non-zero eigenvalues. Hence  $K_{\bf{x}} =U_\Omega \Lambda_{x,s}  U_\Omega^\dagger$ is the reduced  eigenvalue decomposition of   $K_{\bf{x}}$ where  $\Lambda_{x,s} \in  \mathbb{R}^{s \times s}$ is the diagonal matrix of non-zero eigenvalues and $U_\Omega \in \mathbb{C}^{n \times s}$ is the sub-matrix formed by the columns of $U$ corresponding to the non-zero eigenvalues.

%\kern-1em
\subsection{Sensing and Communications to the Fusion Center}%\label{sec:sys:sig}
\begin{figure}
\begin{center}
\begin{footnotesize}
\psset{arrowscale=1}
\psset{unit=0.6cm}
\psset{xunit=1.3,yunit=0.6}
\begin{pspicture}(0,2.0)(5.5,8)
%\psgrid[subgriddiv=0,griddots=10,gridlabels=7pt]
%
\psframe(0.7,2.5)(3.3,4)
\rput(2,3.2){{\small  EH Sensor}}
\rput(0.3,3.3){${ {{x}}_{t}}$}
\rput(4,3.2){${ \sqrt{a_t}{{x}}_{t}}$}
% %
%
% \psframe[linewidth=1pt,fillstyle=solid,fillcolor=lightgray](-1,3)(0,3.5)
% \psframe[linewidth=1pt,fillstyle=solid,fillcolor=lightgray](-2,3)(-1,3.5)
% \psline[linewidth=1pt,linecolor=black]{->}(-4,3.25)(-3,3.25)
% \psline[linewidth=1pt](-3,3.45)(-2,3.45)
% \psline[linewidth=1pt](-3,3.05)(-2,3.05)
% \rput(-2,2.5){{\small Data} {\small Queue}}
% -------------%
\psframe[linewidth=1pt,fillstyle=solid,fillcolor=lightgray](1.75,5)(2.25,5.75)
\psline[linewidth=1pt,linecolor=black]{->}(2,7.0)(2,6.1)
\psframe[linewidth=1pt,fillstyle=solid,fillcolor=lightgray](1.75,5.89)(2.25,6.25)
\psline[linewidth=1pt](1.77,5.75)(1.77,6.5)
\psline[linewidth=1pt](2.22,5.75)(2.22,6.5)
\rput(3.5,6){{\small Battery}}
\rput(2,4.5){\small${{J}_t}$}
\rput(2,7.4){\small${{E}_t}$}
%------------%
% % -------------%
% \psframe[linewidth=1pt,fillstyle=solid,fillcolor=lightgray](1.75,5)(2.25,6.5)
% \psline[linewidth=1pt,linecolor=black]{->}(2,8.75)(2,7.75)
% \psframe[linewidth=1pt,fillstyle=solid,fillcolor=lightgray](1.75,7.0)(2.25,7.25)
% \psline[linewidth=1pt](1.77,6.5)(1.77,7.75)
% \psline[linewidth=1pt](2.22,6.5)(2.22,7.75)
% \rput(3.5,7){{\small Battery}}
% \rput(2,4.5){${\bar{E}_k}$}
% %------------%
% \psframe(-2,3)(-0,3.5)
% \psline[linewidth=1pt](-0.8,3)(-0.8,3.5)
% \psline[linewidth=1pt](-1.6,3)(-1.6,3.5)
% \psline[linewidth=1pt,linecolor=myblue]{->}(-4,3.3)(-3,3.3)
%
%
% \psframe(6.8,4)(9.2,5.5)
% \rput(8,5){${\bf Information }$}
% \rput(8,4.5){${\bf Receiver }$}
% \rput(6.1,4.7){${\bf Y_i}$}
% \rput(9.8,4.7){${\bf { \widehat{S}}_b}$}
% %
% \psframe(6.8,1)(9.2,2.5)
% \rput(8,2){${\bf Energy}$}
% \rput(8,1.5){${\bf  Receiver}$}
% \rput(6.1,1.7){${\bf Y_e}$}
% \rput(9.8,1.7){${\bf  P_e}$}
%\psarcn[linewidth=0.4pt](3 ,3.2 ){1}{45}{-45}
\psarcn[linewidth=0.4pt](4 ,3.2 ){1}{25}{-25}
\psarcn[linewidth=0.4pt](4 ,3.2 ){1.5}{25}{-25}
%\psarcn[linewidth=0.4pt](3 ,3.2 ){3.9}{45}{-45}
\end{pspicture}
\end{footnotesize}
%{\small Energy Harvesting Transmitter } % at transmission time slot $k$
\end{center}
\caption{ Energy Harvesting Sensor}
\label{fig:sys}
%\kern-1.2em
\end{figure}
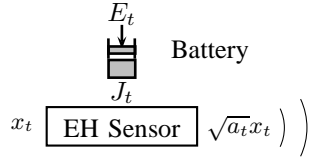

%\kern-0.5em
Motivated by the optimality of uncoded transmission for Gaussian sources over AWGN channels under mean-square error \cite{GastparRimoldiVetterli_2003,gastpar_2008,zhaoChenZhang_2015}  and the high computational and the energy cost of source and channel coding operations, we consider an amplify-and-forward strategy for the sensor similar to \cite{zhaoChenZhang_2015,nourianDeyAhlen_2015,knornDeyAhlenQuevedo_2015}.
% BahceciKhandani_2008,
As illustrated in Fig.~\ref{fig:sys}, at time slot $t$, the sensor measures $x_t$, the unknown signal value at time $t$ and communicates it to the fusion center as follows:
\begin{align}
 {  y_t }=  h_t \sqrt{a_t} x_t + w_t, \quad t=1,\ldots,n
\end{align}
where $h_t \in \mathbb{C}$, $\sqrt{a_t}\in \mathbb{R}$,  $y_t \in \mathbb{C}$ and $w_t \in \mathbb{C}$  denote the channel fading coefficient, the amplification factor adopted by the sensor, the received signal at the fusion center, and  the channel noise respectively.
Here $\mathbf{w} = [ w_1,\ldots, w_n] \in \mathbb{C}^{n \times 1}$  is complex proper zero-mean Gaussian with
$\mathbf{w}  \in \mathbb{C}^{n \times 1}$ $ \sim \mathcal{CN}(0,K_{\mathbf{w}})$, $K_{\mathbf{w}} =\sigma_w^2 I_n$, $\sigma_w^2>0$.

%\kern-1em

%\kern-1em
\subsection{Energy Constraints at the Sensor}
The  average energy used by the sensor  during transmission of $x_t$ can be written as follows \cite{zhaoChenZhang_2015,nourianDeyAhlen_2015,knornDeyAhlenQuevedo_2015} 
\begin{align}\label{eqn:ehCost}
J_{t} \!=\!  \tau \expectation[  || \sqrt{a_k} {{ x}_{t}} ||^2]   \!=\! \tau a_{t}  \sigma_{x_{t}}^2 ,
%\quad t=1,\ldots, n
\end{align}
where the transmit duration is taken as $\tau =1$ in the rest of the paper.  Communications system design under  average power constraints have been considered for a wide range of  scenarios, including amplify-forward strategy design \cite{BahceciKhandani_2008,ShiraziniaDeyCiuonzoRossi_2016} and linear encoder design \cite{LeePetersen_1976}  without energy harvesting constraints.  Here we consider an amplify-forward scenario under EH constraints. 
At each time slot $t$, an energy packet of $E_t$ arrives at the battery.  We consider the off-line scheme, where $E_t$ have arbitrary, but known values, during the time frame $t=1,\ldots,n$ \cite{tutuncuogluYener_2012,YangOzelUlukus_2012,antepliUysalErkal_2011,yangUlukus_2012mac}.
The sensor operates under the following energy neutrality conditions
\begin{align}\label{eqn:en}
\sum_{l=1}^t  J_{l} \leq \sum_{l=1}^t  {{E}_l}, \quad \quad  t = 1,\ldots,n.
\end{align}
where the initial energy at the battery is zero.
These conditions ensure that the energy used at any time does not exceed the available energy.
Here we consider a device with a large enough battery capacity so that no energy packet $E_t$ has to be dropped.

\subsection{Estimation at the Fusion Center}
After receiving $\mathbf{y} =[y_1,\ldots,y_n] \in \mathbb{C}^{n \times 1}$, the fusion center forms the minimum MSE (MMSE) estimate of $\bf{x}$, i.e.  $\hat{\bf{x}}=\expectation[\mathbf{x}|\mathbf{y}]$.
The resulting MMSE can be expressed as\cite[Ch2]{b_andersonMoore_optFiltering}
% \begin{align}\label{eqn:err:long}
% \err(A) =   \expectation [ || \mathbf{x} -  \hat{\mathbf{x}}||^2]
% = \tr[K_{\mathbf{x}} -K_{\mathbf{x}\mathbf{y}} K_{\mathbf{y}}^{-1} K_{\mathbf{x}\mathbf{y}}^\herm]
% \end{align}
% where
% %\begin{align}
%  $ K_{\mathbf{x} \mathbf{y}} = \expectation[\mathbf{x} \mathbf{y}^\herm] = K_{\mathbf{x}} {A}^\herm  $,
%  $ K_{\mathbf{y}} =  A K_{\mathbf{x}} {A}^\herm  + K_{\mathbf{w}}$  and $A=\diag(\sqrt{a_t})\in \mathbb{R}^{n \times n}$.
% %\end{align}
% %
% Hence \eqref{eqn:err:long} can be expressed as
\begin{align}
 \err(A) &\!=\! \tr[K_{\mathbf{x}} -K_{\mathbf{x}\mathbf{y}} K_{\mathbf{y}}^{-1} K_{\mathbf{x}\mathbf{y}}^\herm] %\\
\end{align}
where
\begin{align*}
 \expectation[\mathbf{x} \mathbf{y}^\herm] &=K_{\mathbf{x} \mathbf{y}}  = K_{\mathbf{x}} {A}^\herm H^\herm,\\
 \expectation[\mathbf{y} \mathbf{y}^\herm] &= K_{\mathbf{y}} =  H A K_{\mathbf{x}} {A}^\herm H^\herm + K_{\mathbf{w}},
\end{align*}
 with $H=\diag(h_t)$, $A=\diag(\sqrt{a_t})\in \mathbb{R}^{n \times n}$ and   $\mysnr \doteq {1}/{\sigma_w^2}$.
Hence we have
\begin{align}
  \label{err:invForm}
  \err(A) &= \tr \left[( \Lambda_{x,s}^{-1} + \mysnr U_{\Omega}^\herm \diag(|h_t|^2 a_t)   U_{\Omega})^{-1}\right]. %\\
\end{align}
where \eqref{err:invForm} follows from the Sherman-Morrison-Woodbury identity \cite{HendersonSearle_1981}.
Here the fusion center uses the source and the noise statistics, including the  covariance matrices;  and the amplification factors and the channel gains.  
We note that  the same type of later knowledge  are needed  at the receivers when rate based performance metrics are used \cite{ozelTutuncuogluYangUlukusYener_2011,tutuncuogluYener_2012,YangOzelUlukus_2012,antepliUysalErkal_2011,yangUlukus_2012mac}.
We further discuss these points in Section~\ref{sec:problemStatement}.

We note that by adopting a second-order analysis framework  and using  the optimum {\it{linear}} MMSE filter instead of the MMSE filter at the fusion center,  the above error analysis can be also performed under non-Gaussian statistics.

%
%\kern-1em
\subsection{Problem Statement}\label{sec:problemStatement}
Our goal is to design the optimal transmission strategies in order to minimize the MMSE as follows
\begin{subequations}\label{eqn:opt}
\begin{align}
 \,\, \min_{\,\, \substack{A}} & \quad    \err\left(A\right)   \\
\text{s.t.}
\label{eqn:eh}
& \sum_{l=1}^t  a_{l}  \sigma_{x_{l}}^2 \leq \sum_{l=1}^t  {{E}_l}, \quad \quad  t = 1,\ldots,n-1, \\
\label{eqn:eh:last}
& \sum_{l=1}^n  a_{l}  \sigma_{x_{l}}^2 = E_{tot}, \quad \quad \\
 \label{eqn:pos}
 & a_t \geq 0,  \quad \quad  t = 1,\ldots,n,
\end{align}
\end{subequations}
where  the constraints \eqref{eqn:eh}-\eqref{eqn:eh:last} follow from \eqref{eqn:ehCost}, \eqref{eqn:en} with $ E_{tot} \doteq \sum_{l=1}^n  {{E}_l}$.
Since for any optimum strategy all the available energy should be used, \eqref{eqn:eh:last} is stated as an equality.

Here we consider a scenario where the sensor knows the energy arrivals and the channel gains for a look-ahead window of size $n$, i.e. off-line optimization as investigated for a wide-range of scenarios, including rate-based metrics \cite{tutuncuogluYener_2012,YangOzelUlukus_2012,antepliUysalErkal_2011,yangUlukus_2012mac} and source coding/estimation  \cite{OrhanGunduzErkip_2015, calvofullanaMatamorosAntonHaro_2015,GangulaGunduzGesbert_2015,zhaoChenZhang_2015,nourianDeyAhlen_2015}.
This type of off-line optimization approaches are suitable for  energy harvesting scenarios with dedicated power transfer, for instance  as in \cite{GuoWandYang_2014,duFischioneXiao_2016} where wireless power  transfer is scheduled a priori.
 They  also provide benchmarks  for performance limits of energy harvesting systems and structural guidelines for efficient solutions in the general case.  Examples for this include the online near-optimal scheme of \cite{WangLiu_2013} utilizing the off-line directional water-filling solution of \cite{ozelTutuncuogluYangUlukusYener_2011} and the block transmission scheme of \cite{ayca_isit2016} motivated by the off-line optimal most-majorized power allocation of   \cite[Sec.7]{ozelUlukus_2012}.

We now discuss the convexity properties of the formulation in \eqref{eqn:opt}.
The objective function of \eqref{eqn:opt} is a convex function since $\tr[X^{-1}]$ is convex for $X \succ 0$.
The constraints form convex constraints since they are in the form of linear inequalities and equalities.
Hence  \eqref{eqn:opt} is a convex formulation and the Karush-Kuhn-Tucker (KKT) conditions are necessary and sufficient for optimality  under the assumption of a strictly feasible point.
% %
Optimal solutions can be found using the standard numerical optimization tools, such as SDPT3, SeDuMi and CVX \cite{sdpt3,SeDuMi,cvx}. In Section~\ref{sec:optpolicy} and Section~\ref{sec:lowcomp}, we provide analytical solutions  that reveal the structure of the optimal power allocations for a number of cases and propose low-complexity policies, respectively.  Numerical evaluations are provided in  Section~\ref{sec:num}.

%\kern-1em
\section{Optimal Transmission Policies}\label{sec:optpolicy}

Here we discuss the structure of the solutions for a number of illustrative scenarios.
These results motivate the low-complexity policies proposed in Section~\ref{sec:lowcomp}.

%\kern-1em
\subsection{Uncorrelated Sources}\label{sec:uncorr}
Here we consider the case where the components of  $\mathbf{x}$ are uncorrelated, hence $K_{\mathbf{x}}= \diag(\sigma_{x_t}^2) $, $\sigma_{x_t}^2>0$. The MMSE can then be expressed as follows:
\begin{align} \label{eqn:uncorr}
\err(A) &=\sum_{t=1}^n \frac{ \sigma_{x_t}^2}{1  +  \mysnr |h_t|^2   \sigma_{x_t}^2  a_t}.
\end{align}
% Since \eqref{eqn:opt} is convex formulation,   Karush-Kuhn-Tucker (KKT) conditions are necessary and sufficient for optimality  under the assumption of a strictly feasible point.
% %
%%%%%%
The Lagrangian is given by
\begin{dmath}
 \mathcal{L} =  \sum_{t=1}^n \frac{\sigma_{x_t}^2}{1+\gamma |h_t|^2 \sigma_{x_t}^2  a_t} + \sum_{T=1}^{n-1} \eta_T  W_T
                 + \nu W_n -\sum_{t=1}^{n} \mu_t a_t,
\end{dmath}
where
%$
\begin{align}
 W_k = \sum_{t=1}^{k} \sigma_{x_t}^2  a_t - \sum_{t=1}^{k} E_t, \quad \quad 1 \leq k \leq n
\end{align}
%$, $1 \leq k \leq n$.
Here
 $\eta_T  \in \mathbb{R}$, $ \eta_T \geq 0 $, $1 \leq T \leq n-1$, $\nu \in \mathbb{R}$ and $\mu_t \in \mathbb{R}$, $ \mu_t \geq 0 $, $1 \leq t \leq n$ are the Lagrange multipliers.
 Hence together with the feasibility conditions,  the KKT conditions can be expressed as follows:
 \begin{align}
  \label{eqn:comp1} % \frac{\partial \mathcal{L}}{\partial a_t} &=
   -\frac{ \gamma |h_t|^2 \sigma_{x_t}^4}{(1+ \gamma |h_t|^2 \sigma_{x_t}^2  a_t)^2} &+ \sum_{T=t}^{n-1} \sigma_{x_t}^2 \eta_T +\sigma_{x_t}^2 \nu +\mu_t =0, \quad \forall t  \\
   \label{eqn:comp2}
  \eta_T  W_T &= 0, \quad \quad  T=1,\dots,n-1 \\
  \label{eqn:comp3}
  \mu_t  a_t  &= 0, \quad \quad  t=1,\dots,n
 \end{align}
% with $\eta_T  W_T  = 0$, $1\leq T \leq n-1$ and $\mu_t \sigma_{x_t}^2 a_t = 0$, $t=1\leq t \leq n$.
 %%
 %
Solving the KKT conditions reveals that  the optimal $a_t$ can be expressed as
\begin{align}\label{eqn:uncorr:opta}
%$
 a_t %= \left(\sqrt{\frac{1}{\kappa_t} \frac{1}{\gamma |h_t|^2 \sigma_{x_t}^2}}-\frac{1}{\gamma |h_t|^2 \sigma_{x_t}^2}\right)^{+}
 = \frac{1}{|h_t| \sqrt{\gamma }} \frac{1}{ \sigma_{x_t}^2}  \left(\sqrt{\frac{\sigma_{x_t}^2}{\kappa_t} }-\frac{1}{|h_t|\sqrt{\gamma } }\right)^{+}
%$
 \end{align}
where $c^{+}$ is defined as $c^{+} \doteq \max(0,c)$ and
\begin{align}
%$
\kappa_t \doteq {\sum_{T=t}^{n-1} \eta_T +\nu}
%$
\end{align}
can be interpreted as a time-index dependent threshold, which is a typical property of the EH solutions  \cite{ozelTutuncuogluYangUlukusYener_2011}.  This solution structure dictates that  $x_t$ is sent over the channel with a non-zero power whenever the a priori uncertainty in this component is relatively large, i.e.  ${\sigma_{x_t}^2} >{\kappa_t/(|h_t|^2\gamma) }$. If the a priori uncertainty in this component is relatively small, i.e. this condition is not satisfied,  $a_t$ is chosen as $a_t=0$ and $x_t$ is not sent, hence the energy is saved for future transmissions. We note  that here $1/(|h_t|^2\gamma)$ can be interpreted as the effective channel noise-to-signal ratio, hence for a transmission to occur,  the a priori signal uncertainty should be above the  effective noise-to-signal ratio scaled by $\kappa_t$.

We note that optimal strategies become more generous with energy expenditure as time passes for a static channel, i.e. $h_t=1$.  More precisely,  we obtain the following:

\begin{lemma}
Let $H=I_n$. Let $\tminus \leq \tplus$ denote the ordering of two indices with $1 \leq \tminus,  \tplus  \leq n$. Let $\sigma_{x_\tplus}^2 \geq  \sigma_{x_\tminus}^2 >0$ . Then the following holds:
i) $a_{\tplus} \sigma_{x_\tplus}^2 \geq a_{\tminus} \sigma_{x_\tminus}^2$;
ii) If $a_{\tminus}>0$ , then $a_\tplus>0$.
\end{lemma}

{\it{Proof:}} We note that  $\eta_T \geq 0$, hence we have $\kappa_\tminus \geq \kappa_\tplus$  i.e. $\kappa_t$ is a decreasing function of $t$. Part (i) follows from $\kappa_\tminus \geq \kappa_\tplus$ and $ a_t  \sigma_{x_t}^2  = \frac{1}{\sqrt{\gamma}} (\sqrt{\frac{\sigma_{x_t}^2}{\kappa_t} }-\frac{1}{\sqrt{\gamma} })^{+}$. Part (ii) follows from Part (i)  with $a_{\tminus}>0$.

Part (i) states that if  an energy of $a_t  \sigma_{x_t}^2$ has been used before, one will not use less energy for any subsequent component with higher variance.
Part (ii) states that if a signal component with a given variance has been sent before  (i.e. $a_{\tminus}>0$), all the components with  higher variance (i.e. higher uncertainty) will also be sent over the channel in the future.

We now take  a closer look to the solution structure in the case where the source is white:

\subsubsection{White Sources}\label{sec:iid}
Here $K_{\mathbf{x}}= \sigma_x^2 I_n$ by definition. Under $H=I_n$, the MMSE can be expressed as follows:
\begin{align}\label{eqn:err:iid}
 \err(A) =   \sum_{t=1}^{n} \frac{\sigma_x^2}{1 + \mysnr \sigma_x^2 a_t }. %= \sum_{t=1}^{n} \frac{\sigma_x^2}{1 + \mysnr \sigma_{x_t}^2 a_t } %  \frac{1}{k_{x_i}} + \mysnr a_i
\end{align}
Such sources have been investigated in  \cite{zhaoChenZhang_2015} using the KKT conditions.
Here we adopt an alternative approach and illustrate how optimal strategies can be found by adopting the arguments of \cite{ozelUlukus_2012}.
More precisely, we note the following:
\begin{definition}\cite[Ch.1]{b_marshallOlkin}
  Let $a =[a_1,\ldots,a_n] \in \mathbb{R}^n$ and $ b = [b_1,\dots,b_n]  \in \mathbb{R}^n$. Then $a$ is said to be majorized by $b$  if the following holds:
 \begin{align}
  \sum_{t=1}^k a_{[t]} & \leq   \sum_{t=1}^k b_{[t]}, \quad \quad k=1,\ldots, n-1\\
   \sum_{t=1}^n a_{[t]} &=  \sum_{t=1}^n b_{[t]}
 \end{align}
Here $a_{[t]}$ denotes the components of $a$ in decreasing order, i.e. $a_{[1]} \geq,  \ldots, \geq a_{[n]}$.
This majorization relationship is denoted by  $a \prec b$.
\end{definition}

Majorization can be interpreted as a measure of how balanced the distribution of the components of  vectors are.
In particular, the following relationship holds: $\forall a \in \mathbb{R}$: $\bar{a} \prec a \prec  {\tilde a}$, where $\bar{a}=(S_a/n) [1,\ldots, 1] \in \mathbb{R}^n$ and  ${\tilde a}= [ 0,\ldots, 0, S_a, 0,\ldots, 0] \in \mathbb{R}^n$  has only one non-zero component,  where $S_a =\sum_{t=1}^n a_t$.
Hence, every vector majorizes the vector that has equal components and has the same total sum,  and every vector is majorized by the vector that has only one non-zero component with the same total sum.
The following is of interest:

\begin{definition}\cite[Ch.3]{b_marshallOlkin}
Let us have $\mathcal{S} \subseteq \mathbb{R}^n$ and  $f(.): \mathcal{S} \rightarrow \mathbb{R}$. Then  $f(.)$ is said to be  {\it{Schur-convex}} on $\mathcal{S}$  if $a \prec b$ on $\mathcal{S}$ implies  $f(a) \leq f(b)$.
\end{definition}

\begin{lemma}\cite[Ch.3]{b_marshallOlkin} \label{lem:cvx2schur}
 Let  $\mathcal{S} \subseteq \mathbb{R}$, and $g(.):\mathcal{S} \rightarrow \mathbb{R}$ be  convex. Then $f(a) =\sum_{t=1}^n g(a_t)$ is Schur-convex.
\end{lemma}

By Lemma~\ref{lem:cvx2schur}, \eqref{eqn:err:iid} is Schur-convex since $g(a_t) = \frac{\sigma_x^2}{1 + \mysnr \sigma_x^2 a_t }$ is a convex function of $a_t$, $a_t \geq 0$. Hence an optimal solution is given by $a_t$ that is majorized by all feasible power allocations, i.e. the strategy as balanced as possible, or alternatively as uniformly as possible.
Characterization of such solutions have been studied in relation to maximization of the rate function in \cite{ozelUlukus_2012}:
\begin{lemma}\cite[Thm.3]{ozelUlukus_2012} \label{lem:mostmajorized}
The power allocation that  is majorized by all feasible solutions of \eqref{eqn:eh}, \eqref{eqn:eh:last},   can be characterized as follows: %\eqref{eqn:pos}
\begin{align}\label{eqn:a:uni1}
 \bar{a}_r &= \frac{ \bar{E}_{\tau_k} - \bar{E}_{\tau_{k-1}} }{\tau_k -\tau_{k-1}}, \quad  r = \tau_{k-1}+1,\ldots, \tau_k\\
\label{eqn:a:uni2}
  \tau_k &= \arg \min_{\substack{r \in \{\tau_{k-1}+1,\ldots,\bar{\tau} \} }} \frac{\bar{E}_r -\bar{E}_{\tau_{k-1}}}{r-\tau_{k-1}}, \,\,  k =2,\dots,K
\end{align}
where $1 \leq r \leq n$, $\tau_1 =0$ and $\bar{\tau}=\tau_{K+1}=n$, and $1 \leq K \leq n$ is the number of constant power sections.
\end{lemma}\\
Here we have adopted the notation $\bar{E}_L= \sum_{t=1}^L E_t/\sigma_x^2$,  $a_t=\bar{a}_r$ with $r=t,\forall r,t$  for later notational convenience. Hence we obtain the following:

\begin{corollary}
Let $K_{\mathbf{x}}= \sigma_x^2 I_n$, $H=I_n$. Then \eqref{eqn:a:uni1}-\eqref{eqn:a:uni2} provide an optimal solution for \eqref{eqn:opt}.
\end{corollary}\\
{\it{Proof:}} The result follows from Schur-convexity of \eqref{eqn:err:iid}.

% Due to Schur-convexity of \eqref{eqn:err:iid}, Lemma~\ref{lem:mostmajorized} also provides an optimal strategy for the minimization of the MSE in \eqref{eqn:err:iid}.

In the subsequent sections, we will utilize Lemma~\ref{lem:mostmajorized}  to provide optimal solutions in scenarios even when the source  is not  white.

\subsection{Parameter Estimation}\label{sec:param}
We now consider the scenario where $K_{\mathbf{x}}$ is of rank $1$, hence there is effectively only one random variable to be estimated. We refer to this case as the parameter estimation scenario. In this case, $K_{\mathbf{x}} = U_\Omega \Lambda_{x,1} U_\Omega $ where $U_\Omega \in \mathbb{C}^{n \times 1}$, $\Lambda_{x,1}=P_x$. Let $u_t \in \mathbb{C}$ denote the t$^{th}$ component of $U_\Omega$.  The correlation coefficient between $x_{t_1}$ and $x_{t_2}$ is given by
\begin{align*}
\rho_{t_1 t_2} =\frac{\expectation[x_{t_1} x_{t_2}]}{ \sigma_{x_{t_1}} \sigma_{x_{t_2}} } = \frac{P_x u_{t_1} u_{t_2}^\herm}{(P_x^{1/2} |u_{t_1}|) (P_x^{1/2} |u_{t_2}|)} =\frac{u_{t_1} u_{t_2}^\herm}{|u_{t_1}| |u_{t_2}|}
\end{align*}
% \rho_{t_1 t_2} =\frac{\expectation[x_{t_1} x_{t_2}]}{\expectation[x_{t_1}^2]^{1/2}\expectation[x_{t_2}^2]^{1/2}}
Hence, $|\rho_{t_1 t_2}|=1$, $\forall t_1,t_2$. Hence, when $K_{\mathbf{x}}$ is of rank $1$,  the signal can be said to be fully correlated.
The error  can be expressed as  % in \eqref{err:invForm}
\begin{align}\label{eqn:err:param}
 \err(A) & = \frac{1}{1/P_x+\mysnr \sum_{t=1}^n |h_t|^2 |u_t|^2 a_t} \\
 &=\frac{1}{1+ \mysnr \sum_{t=1}^n  |h_t|^2  \sigma_{x_t}^2 a_t} P_x,
\end{align}
where we have used $|u_t|^2 P_x = \sigma_{x_t}^2$. Optimal solutions can be characterized as follows:
\begin{lemma}\label{lemma:param}
An optimum strategy for \eqref{eqn:opt} for the parameter estimation case is given by the following recursive procedure:\\
 i) Initialization: Let $a_t=0$, $\forall t$. Let $i=1$; $t^*=0$.\\
 ii) Let $S_{i}=[{t^*+1},\ldots,n]$. Let $E_c(t)= \sum_{l={t^*}+1}^t E_l$, $t \in S_{i}$.
 iii) Let ${t^*}=\arg \max_{t \in S_i} |h_t|^2$. Then $a_{t^*}=E_c({t^*})/\sigma_{x_{t^*}}^2$. \\
 iv) If ${t^*} \neq n$, update $i$ as $i=i+1$ and go to Step-ii. Otherwise stop.
\end{lemma}

%{\it{Proof:}}
The proof is given in Section~\ref{sec:pf:param}.
This procedure sends the data in the most favorable time slots, i.e. the time slots with the  highest channel gains, under the energy causality constraints. In particular, in the first iteration,  the time slot with the highest gain   is determined. Let us refer to this time slot as $t_a$.  Hence in the first iteration, a  transmission at $t_a$ with all the energy stored in the battery up to $t_a$  is scheduled (hence no transmission should occur up to $t_a$). In the next  iteration, the time slot with the highest channel gain is found among  the time slots after $t_a$. Let us refer to this time slot as $t_b$, where $t_b \geq t_a$ by construction. The previous procedure is repeated at $t_b$;  all the energy stored in the battery between time slots $t_a$ and $t_b$ is used for  the transmission at $t_b$ and no transmissions should occur in between $t_a$ and $t_b$.  This procedure is repeated until the end of $n$ time steps is reached.  
% where no transmissions should occur between $t^*$ and the high.

We now focus on the static channel case, i.e. $H=I_n$: Since we have $\sum_{t=1}^n  \sigma_{x_t}^2 a_t =E_{tot}$ by \eqref{eqn:eh:last}, looking at \eqref{eqn:err:param} reveals that  any feasible strategy is an optimum strategy including the most uniform strategy given by \eqref{eqn:a:uni1}-\eqref{eqn:a:uni2}. The optimum error value is given by $(1+\mysnr E_{tot})^{-1} P_x$.
This result shows that in the case of a fully correlated source and the static channel, the correlation between the signal values can be used to completely compensate for the unreliability of the EH source as long as the total energy that arrives at the sensor after $n$ time steps stays constant.

%\kern-1em
\subsection{A Lower Bound}
%\kern-0.5em
%
We will now consider a lower bound on the performance. In the upcoming sections, we will utilize this lower bound to prove  the optimality of some  proposed policies.
We consider the following setting:
\begin{subequations} \label{eqn:opt:lower}
\begin{align}
 \,\, \err_{LB} &= \min_{\,\, \substack{A }}  \quad    \err\left(A\right)  \\
 \text{s.t.}
 \label{eqn:eh:last:lower}
 & \sum_{l=1}^n  a_{l}  \sigma_{x_{l}}^2 = E_{tot}  \quad \quad
%% \label{eqn:pos:lower}
%% & a_t \geq 0, , \quad \quad  t = 1,\ldots,n,
%% %
\end{align}
\end{subequations}
subject to \eqref{eqn:pos}.
Compared to \eqref{eqn:opt}, here  the energy causality constraints are ignored and only the total energy constraint is imposed. Hence \eqref{eqn:opt:lower} forms a relaxation of \eqref{eqn:opt} and the optimum value of \eqref{eqn:opt:lower} provides a lower bound for the optimum value of \eqref{eqn:opt}.
We also note that this scenario can be interpreted as fixed battery scenario where a total energy of $E_{tot}$ is available for usage over $n$ time slots.
Such scenarios have been studied in distributed estimation scenarios  under different assumptions \cite{BahceciKhandani_2008,ShiraziniaDeyCiuonzoRossi_2016}.

Let $H=I_n$. To find an analytical expression, we focus on the case where  $\Lambda_{x,s}$ is of the form $\Lambda_{x,s}= \frac{P_x}{s} I_s$, i.e. the non-zero eigenvalues are all equal.
This type of models have been used to represent signal families with a low degree of freedom in various signal applications, for instance  as a sparse signal model in the compressive sensing literature \cite{ayca_unitaryIT2014,TulinoCaireVerduShamai_2013}. 
We obtain the following result for  $\err_{LB}$:
\begin{lemma}\label{lem:LB}
Let $\Lambda_{x,s}= (P_x/s)I_s$, $H=I_n$. Then $a_t = E_{tot}/P_x$, $\forall t$ is an optimum strategy for \eqref{eqn:opt:lower}. The optimal value is given by
%\begin{align}
$
 \err_{LB}= \frac{1}{1+ \mysnr {E_{tot}}/{s} }P_x
 $.
%\end{align}
\end{lemma}

The proof is presented in Section~\ref{sec:pf:LB}. Hence, whenever $a_t = E_{tot}/P_x$ is a feasible allocation for \eqref{eqn:opt}, it is also an optimal strategy. More precisely, we obtain the following result:
\begin{corollary}\label{lem:LB:example}
Let $\Lambda_{x,s}= (P_x/s) I_s$, $H=I_n$. If $\frac{1}{P_x}\sum_{l=1}^t   \sigma_{x_{l}}^2 \leq  \frac{1}{E_{tot}} \sum_{l=1}^t  {{E}_l}$, $\forall t$, then $a_t  =E_{tot}/P_x$ is an optimum strategy for \eqref{eqn:opt} with the optimal value $ \frac{1}{1+ \mysnr {E_{tot}}/{s} }P_x$.
 % todo: The  error is given by
\end{corollary}

A constant energy arrival scenario where the conditions of Corollary~\ref{lem:LB:example} are satisfied is discussed  in Section~\ref{sec:cwss}.

%\kern-1.5em
\subsection{Circularly Wide-Sense Stationary Signals}\label{sec:cwss}
% \kern-0.2em
%
We now focus on the c.w.s.s signals, which constitute a finite dimensional analog of wide-sense stationary signals \cite{neeser_proper_1993,GrayToeplitzReview}.    
By definition, the covariance matrix associated with c.w.s.s. signals  is circulant, i.e. the matrix is determined by its first row as  $[K_{\mathbf{x}}]_{tk}=[K_1]_{\text{mod}_n (k-t)}$, where $K_1 \in \mathbb{C}^{1 \times n}$ is the first row of $K_{\mathbf{x}}$ \cite{neeser_proper_1993,GrayToeplitzReview}. %
Due to the asymptotic equivalence of sequences of circulant and Toeplitz matrices, (which constitute the covariance matrices of wide-sense stationary signals \cite{GrayToeplitzReview}), our  investigations here can be considered as an intermediate step towards understanding limitations imposed by energy harvesting to sensing of wide-sense stationary signals, which is a fundamental signal model  in the fields of communications and signal processing.

Due to circulant structure, we have $\sigma_{x_t}^2 = \sigma_x^2 = P_x/n, \forall t$. 
The unitary matrix $U$ in the EVD of $K_{\mathbf{x}}$ for a circularly wide-sense stationary signal is given by the DFT matrix \cite{neeser_proper_1993,GrayToeplitzReview}.  Let $F^n$ denote the DFT matrix of size $n \times n$, i.e. $[F^n]_{tk} =(1/\sqrt{n}) \exp(-j \frac{2 \pi}{n} (t-1) (k-1))$, $1 \leq t,k \leq n$, where $j=\sqrt{-1}$.
Hence, the reduced EVD of $K_{\mathbf{x}}$ is given by $K_{\mathbf{x}} = F_\Omega^n \Lambda_{x,s} {F_\Omega^n}^\herm $, where $\Lambda_{x,s} = \diag(\lambda_k)\in \mathbb{R}^{s \times s}$ $F_\Omega^n \in \mathbb{C}^{n \times s}$ is the matrix that consists of  $s$ columns of $F^n$ corresponding to non-zero eigenvalues.

{\it{Constant energy arrival scheme with $\Lambda_{x,s}=(P_x/s)I_s$:}
}To gain some insight into the optimal power allocations in the case of c.w.s.s. signals, we consider case with $\Lambda_{x,s}=(P_x/s)I_s$ under constant energy arrival scheme, i.e.  $E_t=E$, $\forall t$.  We observe the following: Due to $\sigma_{x_t}^2 =P_x/n, \forall t$, the conditions of Corollary~\ref{lem:LB:example} are always satisfied for these c.w.s.s. signals.
Hence the lower bound presented in Lemma~\ref{lem:LB} is achieved even under the energy causality constraints in such scenarios.

We now go back to general c.w.s.s. scenario with arbitrary $E_t$ arrivals.  We obtain the following result, which we will utilize later:
\begin{lemma}\label{lem:cwss:rank1perturbation}
Let $H=I_n$.   Let $e_i \in \mathbb{R}^n$, $1 \leq i \leq n$ denote the i$^{th}$ unit vector. Let  the EVD of $K_{\mathbf{x}}$ be given by  $K_{\mathbf{x}} = F^n \Lambda_{x} {F^n}^\herm$ with $\Lambda_{x}= \beta I_{n} + \alpha e_i e_i^\dagger$ with  $ -\beta < \alpha$,  $\beta>0$, $\alpha,\beta \in \mathbb{R}$.
Then  \eqref{eqn:a:uni1}-\eqref{eqn:a:uni2} is an optimal strategy for \eqref{eqn:opt}.
\end{lemma}

The proof is given in Section~\ref{sec:pf:cwss:rank1perturbation}.
The above eigenvalue distribution model covers a number of signal families with appealing interpretations.
We now identify two such cases, i.e. almost white sources and sources with static correlation coefficient.

\subsubsection{Almost White Sources}
When $x_t$ is white, we have $K_{\mathbf{x}}= \sigma_x^2 I_n$. Hence the EVD of $K_{\mathbf{x}}$ is given by  $K_{\mathbf{x}}= U \Lambda_x U^\herm$ with $\Lambda_x = \sigma_x^2 I_n$, where $U$ is an arbitrary unitary matrix since $ U U^\dagger=I_n$ for all unitary matrices. Motivated by this, we refer to the case where  $\Lambda_{x} \propto I_{n}-\epsilon e_j e_j^\dagger$, $0 < \epsilon  < 1$ as an {\it{almost white}} source.

We obtain the following result as a direct corollary to Lemma~\ref{lem:cwss:rank1perturbation}:
Let $H=I_n$. Let $\mathbf{x}$ be almost white with  $K_{\mathbf{x}}=F^n \Lambda_{x} {F^n}^\herm$, $\Lambda_{x} \!=\! I_{n}-\epsilon e_j e_j^\dagger$, $0 < \epsilon < 1$.
Then  \eqref{eqn:a:uni1}-\eqref{eqn:a:uni2} is an optimal strategy for \eqref{eqn:opt}.
This result shows that even when the source is not exactly white but only close to being white as defined above, the most uniform feasible allocation is still an optimal solution.

%\kern-0.1em
\subsubsection{Static Correlation Coefficient}
We now consider the family of signals whose covariance matrix has the following form
\begin{align}
K(\rho) = \frac{P_x}{n}
\begin{bmatrix}
1 & \rho & \ldots  &\rho \\
 \ldots & \ldots &\ldots & \\
\rho & \ldots &   \ldots &  1
\end{bmatrix}
\end{align}
where $K(\rho) \in \mathbb{R}^{n \times n}$, $0 \leq|\rho|\leq 1$, $\rho \in \mathbb{R}$.
Hence, the correlation coefficient between $x_i$ and $x_j$, $i \neq j$ does not depend on $i, j$.
We note that for $K(\rho)$ to be a valid covariance matrix, it should be positive semi-definite, i.e. $K(\rho) \succeq 0$.  Hence, we have $\rho (n-1) +1 \geq 0$, since if this condition were not full-filled,  one would have $v^\herm  K(\rho) v <0$ with $v=[1,\ldots,1] \in \mathbb{R}^{n}$ which contradicts with the requirement  $v^\herm  K(\rho) v \geq 0$, $\forall v \in \mathbb{R}^{n}$  imposed by the definition of positive semi-definite ordering.

We obtain the following result:
\begin{lemma}\label{lem:static}
Let $K_{\mathbf{x}}=K(\rho)$, $H=I_n$. Then,  \eqref{eqn:a:uni1}-\eqref{eqn:a:uni2} is an optimal strategy for \eqref{eqn:opt}.
\end{lemma}

{\it Proof:}
Let $v$ be the first row of $K_{\mathbf{x}}$, i.e.  $v =(P_x/n) [1,\, \rho \ldots \rho] \in \mathbb{C}^n$. Let $z =[z_1,\ldots,z_n]=[\lambda_1,\ldots, \lambda_n] \in \mathbb{R}^n$ be the vector of eigenvalues. The relationship between the eigenvalues and the first row is given by $z= \sqrt{n} F^n v$ \cite{GrayToeplitzReview}. Hence  we obtain ${z}_1 = (P_x/n) (\rho (n-1) +1)$ and ${z}_i = (P_x/n)(1-\rho)$, $2 \leq i \leq n$. Thus, Lemma~\ref{lem:cwss:rank1perturbation} applies. \myQED %todo: check this

\begin{remark}
Regardless of the value of $\rho$, i.e. the level of statistical dependency of the signal components, the strategy that allocates the power as balanced as  possible is an optimal strategy.  
\end{remark}

Although one may expect that as the signal components become more correlated,  strategies that send a low number of signal components with higher power become optimal instead of strategies that allocate power as uniform as possible, Lemma~\ref{lem:static} shows that this may not be always the case and uniform power allocation strategies may continue to be optimal. 
These results complement the other  scenarios where such allocations are found to be optimal, in particular the  i.i.d. sources scenario that follows from the findings of \cite{ozelUlukus_2012} as discussed in Section~\ref{sec:iid} and the sensing of two correlated Gaussian variables studied in a distributed source coding framework in \cite[Prop.3]{GangulaGunduzGesbert_2015}.

\subsubsection{Low-Pass Signals} \label{sec:cwss:lowpass} 
Let $ n/s \in \mathbb{Z}$.
Let us order the eigenvalues of $K_{\mathbf{x}}$ so that $\lambda_k$ denotes the eigenvalue that corresponds to the eigenvector in the k$^{th}$ column of $F^n$, where $F^n$ is as defined above.
Here we consider low-pass signals, i.e. signals for which  $\Omega=\{1,\ldots,s\}$, and $\lambda_1=,\ldots,=\lambda_{s}=P_x/s$, and the rest are zero. Hence we have $K_{\mathbf{x}}=F_\Omega^n \Lambda_{x} {F_\Omega^n}^\herm$, $\Lambda_{x}=(P_x/s) I_s$.

Similar to their deterministic counterparts, given $\sigma_w^2=0$, low-pass c.w.s.s. signals can be recovered from their equidistant samples with zero mean-square error  when the number of samples is larger than $s$, or equivalently the spacing between the samples satisfies $\Delta \leq n/s$   \cite{ayca_unitaryIT2014}.
Motivated by this, we consider communication strategies that send one  out of every  $\Delta =n/s$  samples, i.e. strategies in the form of 
\begin{align}\label{eqn:equdis:a}
   a_t =
\begin{dcases}
   \geq 0  & \text{if}\,\, t= \Delta r + t_d+1, \quad 0 \leq r \leq m-1\\
   0           & \text{otherwise}%\,\,
\end{dcases}
\end{align}
where $m=n/\Delta$ is the number of samples sent, and $t_d \in 0,\ldots, \Delta-1$, the initial delay before sending the first data, is fixed.

We now consider the error associated with the scenario where the sensor only sends these equidistant samples to the fusion center. Let $f_n=\exp(-j \frac{2 \pi}{n})$. Here, $F_\Omega^n$  consists of  the first $s$ columns of $F^n$. Hence equidistantly row sampled $F_\Omega^n$ can be associated with the DFT matrix of size $s \times s$, $F^s$, as follows
\begin{align}
 [F_{\Omega}^n]_{(n/s)r+t_d+1, k+1} &=(1/\sqrt{n}) f_n^{((n/s)r+t_d) k} \\
 &=(1/\sqrt{n}) f_s^{r k} f_n^ {t_d k} \\
 &= \sqrt{s/n} [F^s]_{r+1,k+1} f_n^ {t_d k},
\end{align} where  $ 0\leq k\leq s-1$, $ 0\leq r\leq s-1$. Let $D =\diag(d_k) \in \mathbb{C}^{s \times s}$,  $d_k=f_n^ {t_d k}$.
Let
$
 \bar{a}_r \doteq a_{\Delta r+t_d+1}
$. The error  can be expressed as follows
\begin{align}
 \err(\bar{A}) %& =   \tr[(\frac{s}{P_x} I_s + \mysnr U_{\Omega}^\herm A^\herm A U_{\Omega})^{-1}] \\
           & =   \tr[(\frac{s}{P_x}I_s + \mysnr  \frac{s}{n} D {F^s}^\herm {\bar H}^\herm \bar{A}^\herm \bar{A} {\bar H} {F^s} D)^{-1}], \\
           \label{eqn:equidistant:err:2}
            & =   \tr[(\frac{s}{P_x}I_s +  \mysnr \frac{s}{n} \bar{H}^\dagger \bar{A}^\herm \bar{A}  \bar{H} )^{-1}] \\
           % &=\sum_{r=0}^{s-1} \frac{1}{\frac{s}{P_x}+\frac{s}{n}  \mysnr \bar{a}_r},
            \label{eqn:equidistant:err1}
            & = \sum_{r=0}^{s-1} \frac{1}{\frac{s}{P_x}+\frac{s}{n}  \mysnr \bar{a}_r |\bar{h}_r|^2}\\
            \label{eqn:equidistant:err}
            & = \sum_{r=0}^{s-1} \frac{1}{1+  \mysnr \bar{a}_r \sigma_x^2 |\bar{h}_r|^2}\frac{P_x}{s}
\end{align}
where $\bar{A}=\diag(\sqrt{\bar{a}_r}) \in \mathbb{R}^{s \times s}$ and $\bar{H}=\diag(\bar{h}_r) \in \mathbb{R}^{s \times s}$, $\bar{h}_r=h_{\Delta r+t_d+1}$. Here, \eqref{eqn:equidistant:err:2} follows from the fact that ${F^s}$ and $D$ are  unitary matrices. In \eqref{eqn:equidistant:err}, we have used the fact that $\sigma_x^2 =P_x/n$.
Hence under the equidistant sampling strategy of \eqref{eqn:equdis:a}, \eqref{eqn:opt} can be equivalently expressed as
 \begin{subequations}\label{eqn:opt:lowpass}
\begin{align} \label{eqn:eh:lowpass:err}
 \,\, \min_{\bar{a}_r}  & \quad   \sum_{r=0}^{s-1} \frac{1}{1+  \mysnr \bar{a}_r \sigma_x^2 |\bar{h}_r|^2}    \\ % \frac{P_x}{s}
\text{s.t.}
\label{eqn:eh:lowpass}
& \quad \sum_{r=0}^t  \bar{a}_{l}  \sigma_{x}^2 \leq \sum_{r=0}^t  {{\bar{\bar E}}_r}, \quad \quad  t = 0,\ldots,s-2, %\\
% \label{eqn:eh:last:lowpass}
% & \sum_{l=1}^s  \bar{a}_{l}  \sigma_{x_{l}}^2 = E_{tot}, \quad \quad
\end{align}
\end{subequations}
and  $\sum_{r=0}^{s-1}  \bar{a}_{r}  \sigma_{x}^2 = E_{tot}$ and  $\bar{a}_r \geq 0$.  Here $\bar{\bar E}_r = \sum_{t=t_0}^{\Delta r+t_d+1} E_t$ with $t_0=\max(0,\Delta (r-1)+t_d+2)$.

\begin{remark}
 We observe that   \eqref{eqn:eh:lowpass:err}  and the objective function of Section~\ref{sec:uncorr}, i.e. the error expression in \eqref{eqn:uncorr}, have the same form. Hence with appropriate notational modifications, the water-filling type characterization of optimal power allocations provided by \eqref{eqn:uncorr:opta} also applies to \eqref{eqn:opt:lowpass}.
\end{remark}

We now consider the static channel case, i.e. $H=I_n$.  We obtain the following result:

\begin{lemma}\label{lem:cwss:lowpass} 
Let $H=I_n$, $\Delta=n/s$, $0 \leq  t_d \leq \Delta-1$.
An optimal strategy for \eqref{eqn:opt} under the setting in \eqref{eqn:equdis:a}, i.e. an optimal strategy for \eqref{eqn:opt:lowpass}, is given by \eqref{eqn:a:uni1}-\eqref{eqn:a:uni2} with
$
 \bar{a}_r \doteq a_{\Delta r+t_d+1}
$,
$\bar{E}_r = \sum_{t=1}^{\Delta r+t_d+1} E_t/\sigma_{x}^2$
and $\tau_1 =0$, $\bar{\tau}=\tau_{K+1}=s$, and $1 \leq K \leq s$. % is the number of constant power sections.
\end{lemma}\\
{\it{Proof:}}  By \eqref{eqn:equidistant:err}, under $h_t=1$,  the error can be expressed as $  \err(\bar{A}) = \sum_{r=0}^{s-1} \frac{1}{1+  \mysnr \bar{a}_r \sigma_x^2 }\frac{P_x}{s}$.   Due to Lemma~\ref{lem:cvx2schur}, this is a Schur-convex function. The result then follows from Lemma~\ref{lem:mostmajorized}. \myQED

This strategy allocates the power as uniformly as possible among the $s$ samples sent.
Hence the most balanced feasible power allocation is an optimum strategy for a sampled low-pass c.w.s.s. signal.

The equidistant sampling strategy can also provide optimal solutions for the general scenario of \eqref{eqn:opt} even when the equidistant sampling constraint is not imposed to achievable sensor strategies:

 \begin{corollary}\label{lem:cwss:lowpass:gen}
Let $H=I_n$,  $\Delta=n/s$, $0 \leq  t_d \leq \Delta-1$.
If $\bar{a}_r= E_{tot}/(s \sigma_x^2 )$,
$
 \bar{a}_r \doteq a_{\Delta r+t_d+1}
$
is feasible for \eqref{eqn:opt}, it is an optimal strategy for \eqref{eqn:opt} with an optimum error value of $
 \err_{LB}= \frac{1}{1+ \mysnr {E_{tot}}/{s} }P_x
 $.
 \end{corollary}\\
{\it{Proof:}}
By \eqref{eqn:equidistant:err} and $\bar{a}_r = E_{tot}/(s \sigma_x^2)$, the error can be expressed as
$
                 =  \frac{1}{1+ \mysnr \frac{E_{tot}}{s}} P_x
$.  We observe that the lower bound in Lemma~\ref{lem:LB} is achieved, hence $\bar{a}_r$ is an optimal strategy. \myQED

Hence, if exactly uniform power allocation over equidistant samples is feasible,  sending equidistant samples is an optimal solution for c.w.s.s. signals for the general scenario in \eqref{eqn:opt}.

In general, there may be more than one optimal strategy for \eqref{eqn:opt}. We now provide an example for low-pass c.w.s.s. signals. Let us consider $E_t =E$ $\forall t$ for a static channel.  In this scenario, both of the following power allocations are optimal: i)  $S_{ua}$: uniform power allocation over all the components,  i.e. $ a_t= E_{tot}/(n \sigma_{x}^2) = E/\sigma_{x}^2$, $\forall t$; ii) $S_{ue}$: uniform allocation over the equidistant samples, i.e. $\bar{a}_r=  n E/(s \sigma_{x}^2)$, $\Delta=n/s$,  $t_d=\Delta-1$.  Here optimality of $S_{ua}$ and $S_{ue}$ follow from Corollary~\ref{lem:LB:example} and Corollary~\ref{lem:cwss:lowpass:gen}, respectively.
%We note that both strategies achieve the lower bound in Lemma~\ref{lem:LB}.

%\kern-1em
\section{Low-Complexity Transmission Policies}\label{sec:lowcomp}
%\kern-0.5em
 We now propose a number of heuristic schemes. These schemes provide possibly sub-optimal but nevertheless low-complexity schemes.
 We illustrate the performance of these schemes in Section~\ref{sec:num}.

The objective function in the optimization formulation in \eqref{eqn:opt} includes a matrix inverse which leads to a computationally challenging optimization formulation. Standard numerical optimization tools, such as SDPT3, SeDuMi and CVX \cite{sdpt3,SeDuMi,cvx} convert the problem into a semi-definite programming problem, whose computational complexity is in the order of $O(n^{4.5})$ using an interior-point method \cite{luoMaSoYeZhang_2010}. Due to this high computational complexity, it is of interest to find schemes which  avoid the matrix inverse in \eqref{err:invForm}.
In particular, we consider the following upper bound:
\begin{align}
 \err(A)
  \label{eqn:err:UBII}
& \leq  \sum_{t=1}^n \frac{ \sigma_{x_t}^2}{1  +  \mysnr  |h_t|^2 \sigma_{x_t}^2   a_t} , %\doteq \err_{uc}(A)
\end{align}
where the inequality follows from the fact that the right hand side of \eqref{eqn:err:UBII} is the error of the scheme where  the possible correlation between the signal values are ignored, i.e.  the covariance matrix of $\mathbf{x}$ is assumed to be in the form of $\diag(\sigma_{x_t}^2)$.
  Utilizing the fact that the bound in \eqref{eqn:err:UBII} couples the optimization variables only through a summation,  we propose  a sliding window approach based on the minimization of this upper bound.
Let $1 \leq \lw \leq n \in \mathbb{Z}$ with $n/\lw \in \mathbb{Z}$  be the look-ahead window size.
Let $t_i= (i-1)\lw+1$. At time index $t_i$, $i=1,\ldots,n/\lw$,  the sensor looks ahead $\lw$ time steps and designs the following strategy:
\begin{subequations}\label{eqn:opt:UB}
\begin{align}
 %\,\, A_{UB-lw}^i
 \,\, &\min_{\,\, \substack{a_{t_i},\ldots,a_{t_{i+1}-1} }}  \quad    \sum_{t=t_i}^{t_{i+1}-1} \frac{ \sigma_{x_t}^2}{1  +  \mysnr  |h_t|^2 \sigma_{x_t}^2  a_t}   \\
\text{s.t.}
\label{eqn:eh:UB}
&\,\,\,\,\, \sum_{l=t_i}^t  a_{l}  \sigma_{x_{l}}^2 \leq \sum_{l=t_i}^t  {{E}_l}, \quad \quad  t = t_i,\ldots,t_{i+1}-2, \\
\label{eqn:eh:last:UB}
&\,\, \sum_{l=t_i}^{{t_{i+1}-1}}  a_{l}  \sigma_{x_{l}}^2 = \sum_{l=t_i}^{t_{i+1}-1}  {{E}_l}, \quad \quad %\\
% \label{eqn:pos}
% & a_t \geq 0, , \quad \quad  t = 1,\ldots,n,
%
\end{align}
\end{subequations}
%subject to \eqref{eqn:eh},\eqref{eqn:eh:last}. % and \eqref{eqn:pos}.
The overall strategy $a_t$, $\forall t$ is obtained by solving \eqref{eqn:opt:UB} over $n/\lw$ non-overlapping windows. We observe that any solution found by this approach is a feasible solution for \eqref{eqn:opt}.
We note that using \eqref{eqn:err:UBII} as a performance metric is consistent with the c.w.s.s. signal scenario with the static correlation coefficient, where a balanced power allocation (which is optimal for the uncorrelated case) is an optimal strategy regardless of the correlation  level for a static channel. The performance of \eqref{eqn:opt:UB}  together with a discussion of numerical efficiency is presented in Section~\ref{sec:num}.

We now focus on the case where the non-zero eigenvalues are equal, i.e. $\Lambda_x=({P_x}/{s})I_s$.
We consider the following lower bound:
\begin{lemma}\label{lem:diag:LB}
Let $\Lambda_x=\frac{P_x}{s}I_s$. The following holds:
 \begin{align}\label{eqn:diag:LB}
 \err(A) \geq   \frac{P_x}{s}\left( \sum_{t=1}^n \frac{1}{ 1  +  \mysnr  |h_t|^2 a_t \sigma_{x_t}^2}  +s-n\right)
 \end{align}
\end{lemma}

The proof is given in Section~\ref{sec:pf:diag:LB}.  We observe that this bound also avoids the matrix inverse in the optimization formulation.
 Hence, we propose the transmission strategies that minimizes the right-hand side of \eqref{eqn:diag:LB} as a heuristic strategy as follows
\begin{align}\label{eqn:opt:heur:lower}
\min_{\,\, \substack{a_{t_i},\ldots,a_{t_{i+1}-1} }}  \quad    \sum_{t=t_i}^{t_{i+1}-1} \frac{1}{ 1  +  \mysnr  |h_t|^2 a_t \sigma_{x_t}^2}
\end{align}
subject to \eqref{eqn:eh:UB}, \eqref{eqn:eh:last:UB}.
% 
%  %
 We observe that for the static channel case, by Lemma~\ref{lem:cvx2schur}, the objective function is  Schur-convex and the  optimal strategies are given by the allocation that makes $a_t \sigma_{x_t}^2$ distribution as balanced as  possible. Hence the solutions follow the characterization provided by \eqref{eqn:a:uni1}-\eqref{eqn:a:uni2}  with appropriate notational modifications. In particular for $l_w=n$, we will have  $\bar{a}_r=a_r \sigma_{x_r}^2$ and $\bar{E}_r=\sum_{t=1}^r \bar{E_t}$. % in \eqref{eqn:a:uni1}- \eqref{eqn:a:uni2}.

 We observe that for c.w.s.s. signals (and other signal models with $\sigma_{x_t}^2=\sigma_x^2=P_x/n$), the upper bound given by \eqref{eqn:err:UBII} and the lower bound provided by \eqref{eqn:diag:LB} have the same form, apart from some  scaling factors and  additive terms that do not depend on $a_t$. Hence, the error performance  is bounded as follows:
 \begin{align}\label{eqn:boundsandwich}
   \left( \err_B  +s-n\right) \frac{P_x}{s}  \leq   \err(A)    \leq  \err_B  \frac{P_x}{n},
 \end{align}
 where  $\err_B $ is defined as
$%\begin{align}
\err_B \doteq \sum_{t=1}^n \frac{1}{ 1  +  \mysnr  |h_t|^2  a_t \frac{P_x}{n}}.
$ % \end{align}
For a given $\err_B $, the gap between the upper and lower bounds becomes smaller as the gap between $s$ and $n$ decreases. This is consistent with the fact that as $s$ gets closer to $n$, the signal can be said to be more close to an uncorrelated source. In the limiting case of $s=n$, the bounds are equal as expected, since the inequalities that give rise to both the upper and lower bounds hold with equality in the uncorrelated case.

 For a static channel, minimization of $\err_B$, hence both sides of \eqref{eqn:boundsandwich}, i.e. minimization of both of the bounds provided by \eqref{eqn:err:UBII} and \eqref{eqn:diag:LB}, require us to use the most majorized feasible $a_t \sigma_{x_t}^2$ distribution for c.w.s.s. signals. In Section~\ref{sec:num}, we observe that such solutions can provide performance close to the optimal performance for $\lw=n$. Nevertheless, we note that such strategies are not guaranteed to provide optimal performance. One such scenario is the low-pass c.w.s.s. signals under energy arrivals satisfying the conditions of Corollary~\ref{lem:cwss:lowpass:gen} but not allowing uniform power allocations over all $a_t$.
This point is illustrated in Section~\ref{sec:num}.

%\kern-1em
\section{Numerical Results}\label{sec:num}

\begin{figure}
\begin{center}
\psfrag{YYY}[bc][bc]{\scriptsize Normalized MMSE }
\psfrag{XXX}[Bc][bc]{  \footnotesize{Energy Arrival Rate (p)} }
%\psfrag{DDDATADATA1}{\tiny  $A_{O}$}
%\psfrag{DDDATADATA2}{\tiny $A_G$}
%\psfrag{DDDATADATA3}{\tiny $A_B$}
%\psfrag{DDDATADATA4}{\tiny $A_{U}$-$2$}
%\psfrag{DDDATADATA5}{\tiny $A_{U}$-$4$}
%\psfrag{DDDATADATA6}{\tiny  $A_{U}$-$16$}
\psfrag{DATA1}{\tiny  $A_{O}$}
\psfrag{DATA2}{\tiny $A_G$}
\psfrag{DATA3}{\tiny $A_B$}
\psfrag{DATA4}{\tiny $A_{L}$-$2$}
\psfrag{DATA5}{\tiny $A_{L}$-$4$}
\psfrag{DATA6}{\tiny  $A_{L}$-$16$}
\includegraphics[width=0.75 \linewidth]{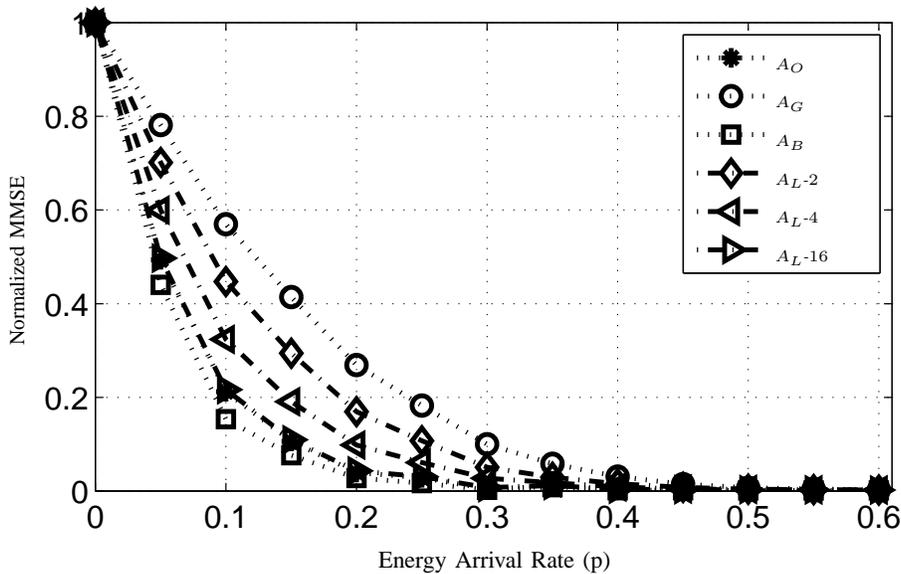} %, height=0.4 \linewidth
\end{center}
\caption{Normalized MMSE versus energy arrival rate,  s = 4.
}
\label{fig:eh:hfading:slow}
\end{figure}

\begin{figure}
\begin{center}
\psfrag{YYY}[bc][bc]{\scriptsize Normalized MMSE}
\psfrag{XXX}[Bc][bc]{  \footnotesize{Energy Arrival Rate (p)} }
%\psfrag{DDDATADATA1}{\tiny  $A_{O}$}
%\psfrag{DDDATADATA2}{\tiny $A_G$}
%\psfrag{DDDATADATA3}{\tiny $A_B$}
%\psfrag{DDDATADATA4}{\tiny $A_{U}$-$2$}
%\psfrag{DDDATADATA5}{\tiny $A_{U}$-$4$}
%\psfrag{DDDATADATA6}{\tiny  $A_{U}$-$16$}
\psfrag{DATA1}{\tiny  $A_{O}$}
\psfrag{DATA2}{\tiny $A_G$}
\psfrag{DATA3}{\tiny $A_B$}
\psfrag{DATA4}{\tiny $A_{L}$-$2$}
\psfrag{DATA5}{\tiny $A_{L}$-$4$}
\psfrag{DATA6}{\tiny  $A_{L}$-$16$}
\includegraphics[width=0.75 \linewidth]{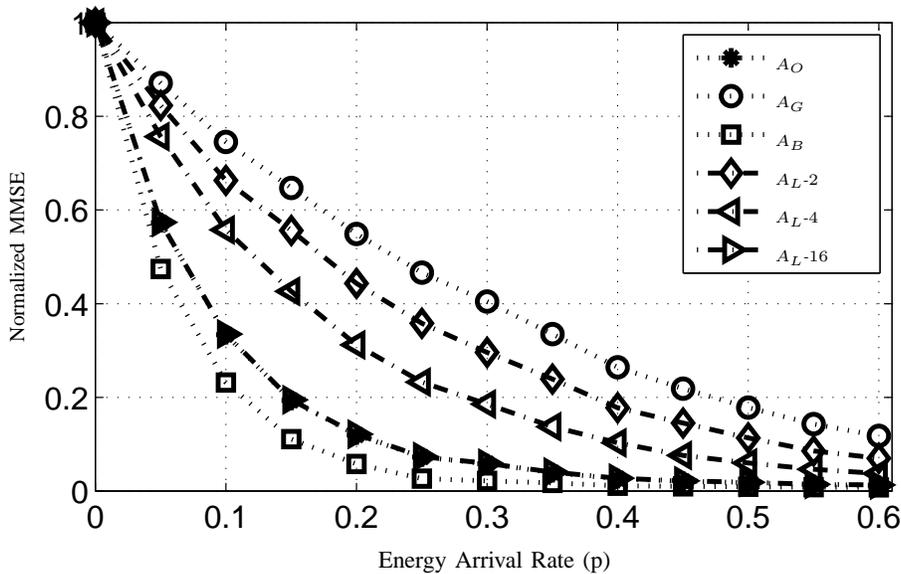}
\end{center}
\caption{Normalized MMSE versus energy arrival rate, s = 14.
}
\label{fig:eh:hfading:shigh}
\end{figure}

We now present the numerical evaluations. 
Let $n\!=\!16$, $s\!=\!4,14$, $P_x \!=\!n$, $\sigma_w^2\!=\! 0.001$, ${\Lambda}_{x,s}=\frac{P_x}{\tr[\Lambda]} \Lambda$, $\Lambda\!=\!\diag(\alpha_k)$, $\alpha_k\!=\!0.7^k, 0 \leq k \leq s-1$. The unitary matrix $U$ is drawn from the uniform (Haar) unitary matrix distribution \cite{Mezzadri_2007} and fixed throughout the experiments unless otherwise stated.
The energy arrivals are generated with $E_t\!=\! \delta_t E_0$, $E_0\!=\!1$, where  $\delta_t$'s are i.i.d. Bernoulli with probability of success $p$, $0 \leq p \leq 1$.
We generate $h_t$ as i.i.d. complex proper Gaussian with $h_t \in \mathbb{C}$, $ h_t \sim \mathcal{CN}(0,1)$.
The average error over $N_{sim}\!=\!500$ realizations are reported. The error is normalized as $\err/P_x$. % $\prop(\delta_t=1)=p$
 The solutions  provided by  \eqref{eqn:opt}, \eqref{eqn:opt:UB} and \eqref{eqn:opt:heur:lower} are denoted by $A_O$, $A_{U}$-$\lw$, $A_{L}$-$\lw$, respectively. The greedy approach where the energy is spent as soon as it arrives is denoted by $A_{G}$ and the lower bound in \eqref{eqn:opt:lower} that ignores the energy neutrality conditions  is denoted by $A_B$.

The error versus energy arrival rate curves  are presented  in Fig.~\ref{fig:eh:hfading:slow} and Fig.~\ref{fig:eh:hfading:shigh}, for $s=4$ and $s=16$, respectively.
As expected, due to the low degree of freedom of the signal and the possible high correlation between the signal values, it is possible to obtain lower error values in Fig.~\ref{fig:eh:hfading:slow}.
In Fig.~\ref{fig:eh:hfading:slow},  the gap between the performance of the optimal and the sub-optimal schemes including the greedy scheme is relatively small compared to Fig.~\ref{fig:eh:hfading:shigh}.
This is consistent with the low degree of freedom of the signal, i.e. $s=4$,  and the relative insensitivity of the performance to the energy allocation  as suggested by the limiting case of parameter estimation scenario and the compressive sensing results \cite{ayca_unitaryIT2014}.

\begin{figure}
\begin{center}
\psfrag{YYY}[bc][bc]{\scriptsize Normalized MMSE }
\psfrag{XXX}[Bc][bc]{  \footnotesize{Energy Arrival Rate (p)} }
%\psfrag{DDDATADATA1}{\tiny  $A_{O}$}
%\psfrag{DDDATADATA2}{\tiny $A_G$}
%\psfrag{DDDATADATA3}{\tiny $A_B$}
%\psfrag{DDDATADATA4}{\tiny $A_{L}$-$2$}
%\psfrag{DDDATADATA5}{\tiny $A_{L}$-$4$}
%\psfrag{DDDATADATA6}{\tiny  $A_{L}$-$16$}
\psfrag{DATA1}{\tiny  $A_{O}$}
\psfrag{DATA2}{\tiny $A_G$}
\psfrag{DATA3}{\tiny $A_B$}
\psfrag{DATA4}{\tiny $A_{L}$-$2$}
\psfrag{DATA5}{\tiny $A_{L}$-$4$}
\psfrag{DATA6}{\tiny  $A_{L}$-$16$}
\includegraphics[width=0.75 \linewidth]{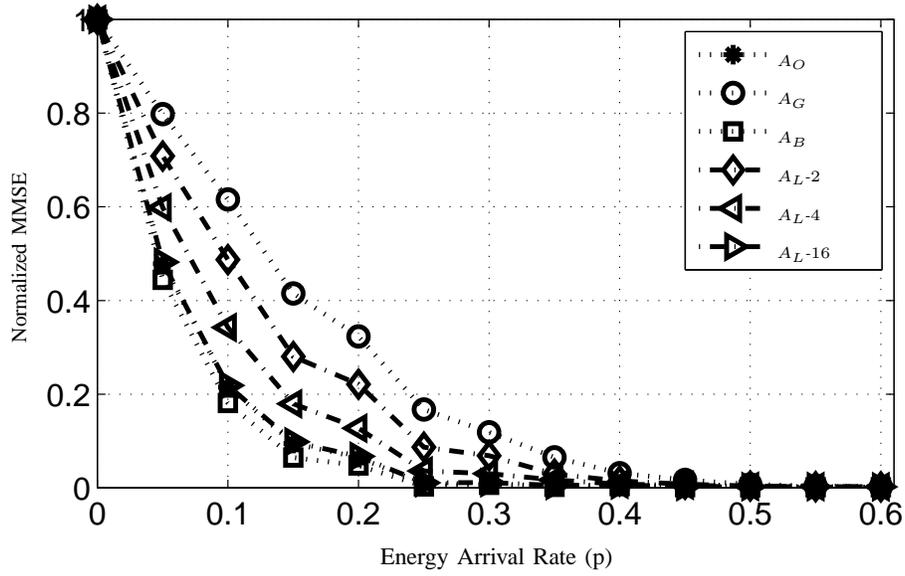}
\end{center}
\caption{Normalized MMSE versus energy arrival rate, static channel, c.w.s.s. signal, s = 4.
}
\label{fig:eh:hfading:slow:DFT}
\end{figure}
\begin{figure}
\begin{center}
\psfrag{YYY}[bc][bc]{\scriptsize Normalized MMSE}
\psfrag{XXX}[Bc][bc]{  \footnotesize{Energy Arrival Rate (p)} }
%\psfrag{DDDATADATA1}{\tiny  $A_{O}$}
%\psfrag{DDDATADATA2}{\tiny $A_G$}
%\psfrag{DDDATADATA3}{\tiny $A_B$}
%\psfrag{DDDATADATA4}{\tiny $A_{L}$-$2$}
%\psfrag{DDDATADATA5}{\tiny $A_{L}$-$4$}
%\psfrag{DDDATADATA6}{\tiny  $A_{L}$-$16$}
\psfrag{DATA1}{\tiny  $A_{O}$}
\psfrag{DATA2}{\tiny $A_G$}
\psfrag{DATA3}{\tiny $A_B$}
\psfrag{DATA4}{\tiny $A_{L}$-$2$}
\psfrag{DATA5}{\tiny $A_{L}$-$4$}
\psfrag{DATA6}{\tiny  $A_{L}$-$16$}
\includegraphics[width=0.75 \linewidth]{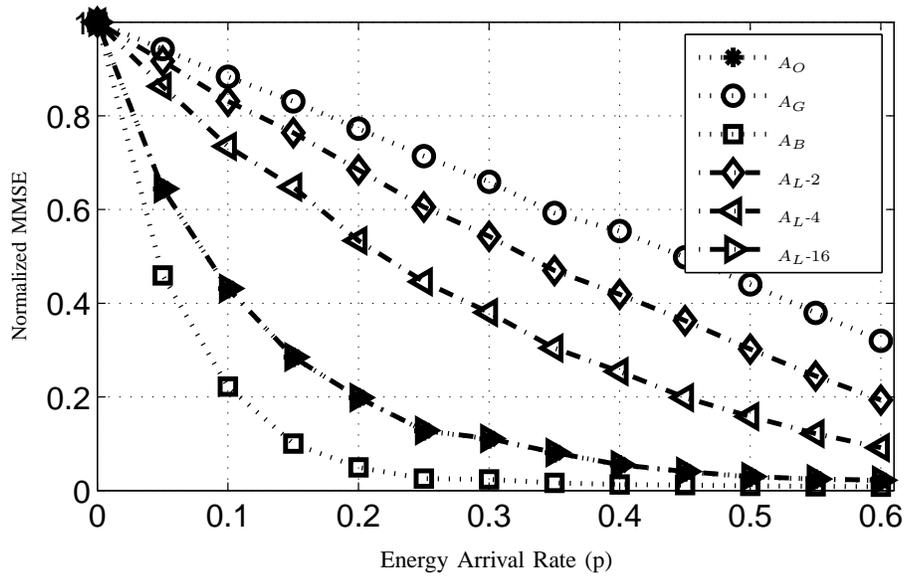}
\end{center}
\caption{Normalized MMSE versus energy arrival rate, static channel, c.w.s.s. signal, s = 14.
}
\label{fig:eh:hfading:shigh:DFT}
\end{figure}

In both scenarios, the low-complexity scheme with the look-ahead window of $\lw\!=\!n$,  $A_{U}$-$n$, is remarkably successful so that  the performance of $A_O$ and $A_{U}$-$n$ are almost indistinguishable from each other in the plots.
In the case of Fig.~\ref{fig:eh:hfading:slow}, this is again consistent with the relative insensitivity of the performance to the energy allocation for a signal with low degree of freedom as suggested by the parameter estimation scenario  and the fact that the correlation may have limited effect on the optimal strategies, as illustrated by the static correlation coefficient case. In the case of Fig.~\ref{fig:eh:hfading:shigh}, the close performance of $A_O$ and $A_{U}$-$\lw$ is supported by the relative closeness of the source to an uncorrelated source due to the relatively high degree of freedom provided by $s=14$.
We note that despite this close average performance, the performance gap may be relatively significant for some  realizations, and the power allocations provided by $A_O$ and $A_{U}$-$n$  may be different.
We illustrate these points later in this section.

The error versus energy arrival rate curves for  the c.w.s.s. scenarios are presented in Fig.~\ref{fig:eh:hfading:slow:DFT} and Fig.~\ref{fig:eh:hfading:shigh:DFT}, for $s=4$ and $s=16$, respectively. Here we have considered the flat eigenvalue distribution scenario with ${\Lambda}_{x,s}=\frac{P_x}{\tr[\Lambda]} \Lambda$, $\Lambda\!=\!\diag(\alpha_k)$, $\alpha_k\!=\!1, 0 \leq k \leq s-1$ so that $A_{L}$-$\lw$ applies. The  performance of the low-complexity policies $A_{U}$-$\lw$ and $A_{L}$-$\lw$ are very close, hence we only present the performance of $A_{L}$-$\lw$ to avoid clutter in the figures.  We observe that again with small $s$, it is possible to obtain lower error values. Similar to $A_{U}$-$n$, the performance of $A_{L}$-$n$ is close to  the performance of optimal policies.

\begin{figure}
\begin{center}
\psfrag{YYY}[bc][bc]{\scriptsize Performance Gap ($e_{G}$)}
\psfrag{XXX}[Bc][bc]{  \footnotesize{Energy Arrival Rate (p)} }
\psfrag{DDDATA1}{\tiny  $s=4$}
%\psfrag{DDDATADATA2}{\tiny $s=14$}
\includegraphics[width=0.75 \linewidth]{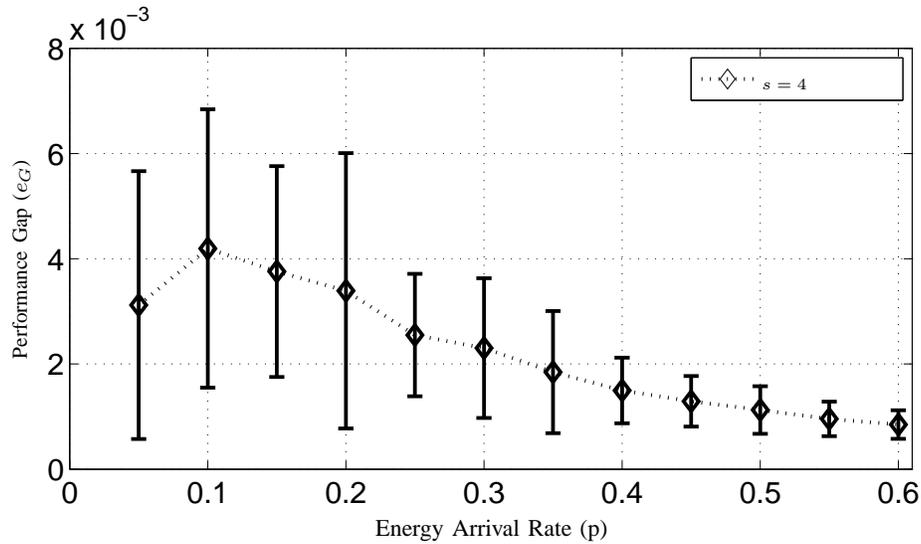}
\end{center}
\caption{Performance gap between $A_{U}$-$n$ and $A_{O}$ versus energy arrival rate, $s=4$.
}
\label{fig:comp1}
\end{figure}

\begin{figure}
\begin{center}
\psfrag{YYY}[bc][bc]{\scriptsize Performance Gap ($e_{G}$)}
\psfrag{XXX}[Bc][bc]{  \footnotesize{Energy Arrival Rate (p)} }
\psfrag{DDDATA1}{\tiny  $s=14$}
%\psfrag{DDDATADATA2}{\tiny $s=14$}
\includegraphics[width=0.75 \linewidth]{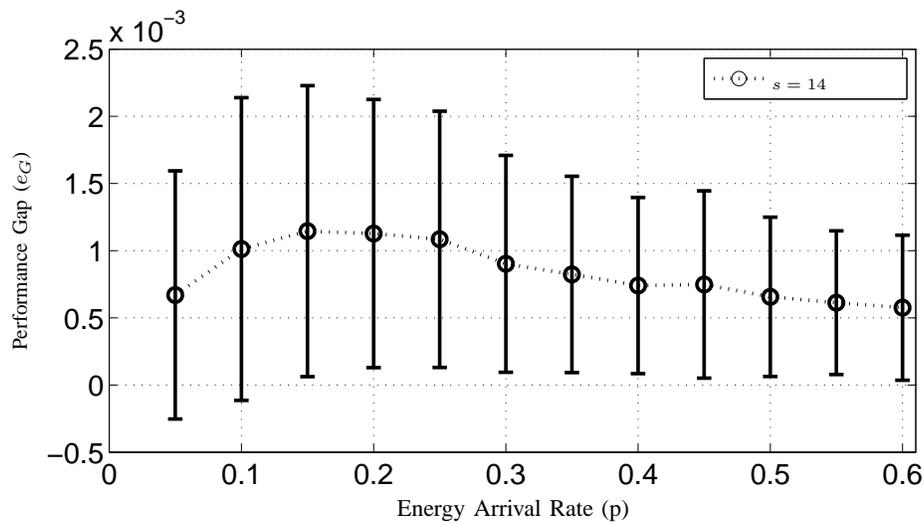}
\end{center}
\caption{Performance gap between $A_{U}$-$n$ and $A_{O}$ versus energy arrival rate, $s=14$.
}
\label{fig:comp2}
\end{figure}

 We now take a closer look at the performance gap between the optimal policies and the low-complexity policies.
 Let $e_O$  and $e_{U}^n$  denote  the error associated with  $A_O$ and $A_{U}$-$n$, respectively. Let us denote the error gap as $e_{G} = \text{$e_{U}^n$} - e_O$, for a given EH realization. We denote the average and the standard deviation of $e_{g}$ over $N_{sim}$ different simulation realizations in Fig.~\ref{fig:comp1}  and  Fig.~\ref{fig:comp2},  for the  setting in Fig.~\ref{fig:eh:hfading:slow} and the setting in Fig.~\ref{fig:eh:hfading:shigh} respectively. Here the deviation is presented with an error bar with a length of one  standard deviation on the mean values. We note that, consistent with the presentation of error values in Fig.~\ref{fig:eh:hfading:slow} and Fig.~\ref{fig:eh:hfading:shigh}, Fig.~\ref{fig:comp1}  and  Fig.~\ref{fig:comp2} report the gaps on the normalized error values i.e. $\err/P_x$.
 We observe that both the mean and the standard deviation are small, illustrating that for most of the EH realizations low-complexity policies $A_{U}$-$n$ provide performance close to the optimal. We note that the  gap between the performance of $A_O$ and that of $A_{L}$-$n$ shows similar behaviour, which is not presented here to avoid repetition in the figures.

Despite this close average performance,  the power allocations provided by $A_O$ and $A_{U}$-$n$ may be different. One such scenario is the low-pass c.w.s.s. signals under energy arrivals satisfying the conditions of Corollary~\ref{lem:cwss:lowpass:gen} but not allowing uniform power allocations over all $a_t$. We now provide a scenario that illustrates this. Let $x$ be a low-pass c.w.s.s. with $s=4$ with ${\Lambda}_{x,s}=\frac{P_x}{\tr[\Lambda]} \Lambda$, $\Lambda\!=\!\diag(\alpha_k)$, $\alpha_k\!=\!1, 0 \leq k \leq s-1$. Let $E_{t}=1$, for $t=4 k$, $1 \leq k \leq s$ and zero otherwise. By Corollary~\ref{lem:cwss:lowpass:gen}, the uniform allocation over equidistant samples, i.e. $a_{t}=1$ for $t=4 k$, $1 \leq k \leq s$ and zero otherwise is an optimal strategy. On the other hand, $A_{U}$-$n$ provides the most majorized strategy which is given by $a_{t}=0.25$ for $4 \leq t \leq n-1$,  $a_{n}=1$ and zero otherwise. These allocations result in a normalized error of approximately $ 9.99 \times 10^{-4}$ and  $1.16 \times 10^{-3}$   for  $A_O$ and $A_{U}$-$n$, respectively.

We now illustrate the numerical efficiency  of the sub-optimal approaches.
The average computational time  of $A_U$ together with that of $A_O$ is provided in Table~\ref{table:num} for $n=s$, $p=0.3$. The optimization problem solved by $A_L$  has the same structure as the one for $A_U$, hence it leads to similar values and is omitted. In Table~\ref{table:num}, the values are normalized  with the value for $A_O$ with $n=16$.
We observe that although the computational time increases for all approaches with increasing $n$, this effect is most prominent for the approach that directly solves the optimization problem in \eqref{eqn:opt} i.e. $A_O$. Comparing the computational time for  $A_U$-$\lw$ for different values of $\lw$, the total time is observed to be higher with small $\lw$ compared to $\lw=n$. This is due to the sliding window approach which requires $n/\lw$ calls to the optimization procedure.
Although for $n=16$  the run-time of  $A_U$-$2$ is higher than that of $A_O$, $A_U$ becomes the most numerically efficient approach for all $\lw$ with increasing $n$. We observe that as $n$ increases, the gap between the computational  time values for the direct optimization approach $A_O$ and the sub-optimal approach of $A_{U}$-$n$ increases significantly. Together with the close performance of $A_{U}$-$n$ to $A_O$, this supports the usage of $A_{U}$-$n$ as a possibly sub-optimal but nevertheless a numerically efficient approach.

% , i.e. $0.79$ secs
%

\begin{table}%[h]
\renewcommand{\arraystretch}{1.3}
\caption{\\  Normalized Average Computational Time }
\label{table:num}
\kern-1em
\begin{center}
\begin{footnotesize}
\begin{tabular}{|l|c|c|c|c|}
\hline
  & $A_O$ &   $A_{U}$-$2$  & $A_{U}$-$n/2$  & $A_{U}$-$n$  \\
\hline
{\footnotesize $n=16$ }   & 1 &  2.64 & 0.70   & 0.41\\
 \hline
{\footnotesize $n=32$ }   & 5.42 &  5.19 & 0.82 & 0.55  \\
 \hline
 {\footnotesize $n=64$ }   & 80.50 &  10.32 &  1.11 & 0.85\\
 \hline
\end{tabular}
 \end{footnotesize}
\end{center}
\kern-1em
\end{table}

%\kern-0.4em
\section{Conclusions}\label{sec:conc}

We have focused on the remote estimation of a time-correlated signal using an EH sensor.
We have considered the problem of optimal power allocation at the sensor under energy causality constraints in order to minimize the MSE at the fusion center. 
Contrary to traditional line of work, the correlation between the signal values was an important aspect of our formulation. 
We have provided structural results for the optimal power allocation strategies for a number of scenarios.
In the case circularly wide sense stationary signals, we have showed that the optimal strategy can be characterized as a water-filling type solution for  sampled low-pass signals for a fading channel.
We have showed that the most majorized power allocation strategy, i.e. the power allocation as balanced as possible,  is optimal regardless of the degree of correlation in the case of c.w.s.s. signals with a static correlation coefficient and  in the case of sampled low-pass c.w.s.s. signals for a static channel.
These results provided important insights into remote estimation of correlated signals under EH constraints that cannot be obtained by considering uncorrelated signals. 
Due to asymptotic equivalence of covariance of matrices of c.w.s.s. signals  and Toeplitz matrices, these investigations can be seen as an intermediate step towards understanding the limitations imposed by energy harvesting constraints on sensing of wide-sense stationary signals.

We  have  proposed low-complexity policies for the general case based on upper and lower bounds on the mean-square error. Numerical evaluations have illustrated the performance of low-complexity and optimal policies.  These results demonstrated the effect of the energy harvesting constraints and the trade-offs associated with the degree of freedom of the unknown signal.
 The close-to-optimal performance of  the low-complexity approaches  with the full look-ahead window and the improvements offered by these approaches in terms of computational time,  support the usage of these low-complexity policies as promising, possibly sub-optimal but nevertheless numerically efficient strategies.

%\kern-0.5em
\section{Appendix}
%\kern-0.3em

\subsection{Proof of Lemma~\ref{lemma:param}}\label{sec:pf:param}
We note that in the parameter estimation case, minimizing $\err(A)$ is equivalent to maximizing the sum  $\sum_{t=1}^n  |h_t|^2  \sigma_{x_t}^2 a_t$. We first consider the case without the energy causality constraints, i.e.
 \begin{align}\label{eqn:opt:param}
\max_{q_t} \sum_{t=1}^n  |h_t|^2  q_t
 \end{align}
 subject to $\sum_{t=1}^n  q_t =E_{tot}$,  $q_t \geq 0$ where $q_t = \sigma_{x_t}^2 a_t$.
The optimal strategy for \eqref{eqn:opt:param} is given as follows: $q_{{t^*}}=E_{tot}$, ${t^*}=\arg \max_{1 \leq t \leq n} |h_t|^2$, and $q_t=0$, if $t \neq {t^*}$. Hence the optimal strategy is in the form of transmission with all the available energy  in the slot with the highest gain.  Optimality of this strategy can be seen, for instance, by observing that any other strategy will achieve a smaller objective function since $|h_t|^2 \leq |h_{t^*}|^2$ for $t \neq t^*$. We note that if different time slots have the same maximum channel gain, i.e $|h_{{t^*}}|^2=|h_{t_1}|^2=|h_{t_2}|^2$, $t_1 \neq t_2$, the energy can be allocated arbitrarily between these time slots.

We now go back to the original setting of Lemma~\ref{lemma:param} with the energy causality constraints. We observe that at the first iteration, the procedure gives the optimal possible allocation for the energy allocation up to time $t^*$. We also observe that one cannot improve the objective function by saving some of this energy for future transmissions since $|h_t|^2 \leq |h_{t^*}|^2$ for $t >t^*$. Similar to the previous case, if we have $|h_{{t^*}}|^2=|h_{t_1}|^2=|h_{t_2}|^2$, $t_1 \neq t_2$, the energy saved up to $t=\max(t_1,t_2)$ can be allocated to the transmissions at $t_1$ and $t_2$ in an arbitrary  manner (under the condition energy causality constraints are not violated) without any change in the objective function. Thus, at any iteration $i$,  Step-iii of Lemma~\ref{lemma:param} provides an optimal allocation up to $t^*$ at that iteration.
Hence the procedure given in  Lemma~\ref{lemma:param} provides an optimal strategy.

\subsection{Proof of Lemma~\ref{lem:LB}}\label{sec:pf:LB}

Let  $R_A = \frac{P_x}{s} U_{\Omega}^\herm \diag( a_t)  U_{\Omega}$.
We observe that
\begin{align}
 \tr[R_A]  &=  \frac{P_x}{s} \tr [ U_{\Omega}^\herm \diag( a_t) U_{\Omega}]\\
      \label{eqn:lower:tr1}
           &=  \frac{P_x}{s} \tr[\diag( a_t) U_{\Omega} U_{\Omega}^\herm] \\
%  \end{align}
%            and hence
% \begin{align}
   \label{eqn:lower:tr2}
           &=  \tr[\diag( a_t) K_{\mathbf{x}}]\\
           &= \sum_{t=1}^n  a_t \sigma_{x_t}^2 \\
               \label{eqn:lower:tr}
           & =  E_{tot}
\end{align}
where  we have used    $\tr[A B] =\tr[B A]$ for matrices with appropriate dimensions in \eqref{eqn:lower:tr1},  $K_{\mathbf{x}} =  \frac{P_x}{s}  U_{\Omega} U_{\Omega}^\herm$ in \eqref{eqn:lower:tr2} and  \eqref{eqn:eh:last:lower} in \eqref{eqn:lower:tr}. %  is equivalent to the condition $\tr[R_A] = \sum_i \lambda_i(R_A) = E_{tot}$.
We now consider the error expression
\begin{align}
 \err(A)  & =   \tr \left[(\frac{s}{P_x} I_s  +  \mysnr U_{\Omega}^\herm \diag( a_t) U_{\Omega})^{-1}\right],\\
%  \label{eqn:lower:1}
%           & =   \tr \left[( \frac{s}{P_x} I_s  +  \mysnr \diag( a_i) U_{\Omega} U_{\Omega}^\herm  )^{-1}\right] \\
  \label{eqn:lower:2}
           &= \sum_{i=1}^s \frac{1}{1+ \mysnr \lambda_i(R_A ) } \frac{P_x}{s},\\
             \label{eqn:lower:3}
           &\geq \sum_{i=1}^s \frac{1}{1+ \mysnr \frac{\tr[R_A]}{s} } \frac{P_x}{s},
\end{align}
where $\lambda_i(R_A)$ denotes the eigenvalues of $R_A$.
Since \eqref{eqn:lower:2} is a Schur-convex function of $\lambda_i(R_A)$, it is lower bounded  by \eqref{eqn:lower:3} which is the error associated with   a uniform eigenvalue distribution for $R_A$, i.e.  $\lambda_i (R_A)= \tr[R_A]/s = E_{tot}/s$, $i=1,\ldots,s$.
This lower bound in \eqref{eqn:lower:3} is achievable by  choosing $a_t  =E_{tot}/P_x$. In particular, this choice of $a_t$ results in $\lambda_i (R_A)= E_{tot}/s$, since  $ R_A=({P_x}/{s})U_{\Omega}^\herm \diag( E_{tot}/P_x)  U_{\Omega} =  (E_{tot}/s) I_s$ where we have used $U_{\Omega}^\herm  U_{\Omega} = I_s$. % \myqed

% $ ({P_x}/{s})U_{\Omega}^\herm \diag( a_t)  U_{\Omega} =  (E_{tot}/P_x) ({P_x}/{s}) U_{\Omega}^\herm U_{\Omega} = (E_{tot}/s) I_s$.

%%%%%%%%%%%%%%%%%%%%%%%%%%%%%%%%%%%%%%%%%%%%%%%%%%%%%%
%\kern-0.5em
\subsection{Proof of Lemma~\ref{lem:cwss:rank1perturbation}}\label{sec:pf:cwss:rank1perturbation}
%\kern-0.4em
We first recall that a function  of $n$ variables whose value does not change for any permutation of the input is called symmetric \cite{b_marshallOlkin}.
We rewrite  $\err(A)$ to show it is a symmetric function of $a_1,\ldots,a_n$ as follows
\begin{align}
  \label{eqn:pert:0}
 \err(A)  & =   \tr \left[ (\bar{\beta} I_n + \bar{\alpha} e_j e_j^\dagger  +  \mysnr {F^n}^\herm \diag( a_t) F^n)^{-1}\right],\\
  \label{eqn:pert:1}
         & =   \tr \left[R^{-1} - \frac{ R^{-1} \bar{\alpha} e_j e_j^\dagger R^{-1} }{ 1+\bar{\alpha} e_j^\dagger R^{-1}  e_j} \right], \\
     \label{eqn:pert:12}
         & =   \tr \left[R^{-1}\right] - \frac{ \bar{\alpha} e_j^\dagger R^{-2}  e_j }{ 1+\bar{\alpha} e_j^\dagger R^{-1}  e_j} ,
%            \label{eqn:pert:2}
%           & = \sum_{t=1}^n \theta_t-  \frac{\bar{\alpha}}{1+\bar{\alpha} \frac{1}{n} \sum_{t=1}^n \theta_t}  \frac{1}{n} \sum_{t=1}^n \theta_t^2
\end{align}
 where $\bar{\alpha} =  1/(\alpha+\beta)-1/\beta $, $\bar{\beta} =1/\beta >0$ and
 \begin{align*}
 R = \bar{\beta} I_n +  \mysnr {F^n}^\herm \diag( a_t) F^n = {F^n}^\herm   \diag(\bar{\beta} + \mysnr a_t) F^n.
 \end{align*}
  Here  \eqref{eqn:pert:1} follows from  the Sherman-Morrison-Woodbury identity  with $1+\bar{\alpha} e_j^\dagger R^{-1}  e_j \neq 0$ \cite{HendersonSearle_1981} and      \eqref{eqn:pert:12} follows from $\tr[AB]=\tr[BA]$ for matrices with appropriate dimensions.
  Let $\theta_t= 1/(\bar{\beta} + \mysnr a_t)$, hence $R={F^n}^\herm   \diag(1/\theta_t) F^n$ and  $R^{-1} = {F^n}^\herm   \diag(\theta_t) F^n$. We have
  \begin{align}
   [R^{-1}]_{ii} = e_i^\dagger R^{-1}  e_i =\sum_{t=1}^n \theta_t |[F^n]_{it}|^2 = \frac{1}{n} \sum_{t=1}^n \theta_t
  \end{align}
 and similarly $[R^{-2}]_{ii}=(1/n) \sum_{t=1}^n \theta_t^2 $. Hence we obtain
 \begin{align}
  \label{eqn:pert:2}
 \err(A)       & = \sum_{t=1}^n \theta_t-  \frac{\bar{\alpha}}{1+\bar{\alpha} \frac{1}{n} \sum_{t=1}^n \theta_t}  \frac{1}{n} \sum_{t=1}^n \theta_t^2.
\end{align}
 Here \eqref{eqn:pert:2} reveals that $\err(A)$ is a symmetric function of $a_1,\ldots,a_n$. Since $\err(A)$ is also a convex function of $a_t$, (due to, for instance, \eqref{eqn:pert:0} and the fact that $\tr[X^{-1}]$ is convex for $X \succ 0$) $\err(A)$ is Schur-convex by \cite[Ch.3-Prop.C2]{b_marshallOlkin}. The result follows from Lemma~\ref{lem:mostmajorized}.

\subsection{Proof of Lemma~\ref{lem:diag:LB}}\label{sec:pf:diag:LB}
%Let $\lambda=\frac{P_x}{s}$.
Let $d_t=|h_t|^2 a_t$ $\forall t$.  We have
\begin{align}
 \err(A)          & = \frac{P_x}{s} \left(\tr \left[( I_s  +  \mysnr \frac{P_x}{s} U_{\Omega}^\herm \diag( d_t) U_{\Omega})^{-1}\right] \right) \\
                \label{eqn:heur:lower0}
                 & = \frac{P_x}{s} \left( \tr \left[( I_n  +  \mysnr  \diag( d_t) U_{\Omega}  \frac{P_x}{s} U_{\Omega}^\herm )^{-1}\right] +s-n\right) \\
                             & =  \frac{P_x}{s}  \left( \tr \left[( I_n  +  \mysnr  \diag(d_t) K_{\mathbf{x}} )^{-1}\right] +s-n\right) \\
                             \label{eqn:heur:lower2}
                            &  \geq    \frac{P_x}{s}\left( \sum_{t=1}^n \frac{1}{ 1  +  \mysnr  |h_t|^2  a_t \sigma_{x_t}^2}  +s-n\right)
\end{align}
The equality in \eqref{eqn:heur:lower0} follows from the equivalence of the non-zero eigenvalues of the matrix products $A B$ and $B A$; see, for instance, \cite[Ch9-A.1.a]{b_marshallOlkin}. The inequality in  \eqref{eqn:heur:lower2} is due to the fact that for a Hermitian matrix $R \in \mathbb{C}^{s \times s}$, $\tr[R^{-1}] \geq \tr [\diag([R]_{ii})^{-1}]$,  which, for instance, follows from Lemma~\ref{lem:cvx2schur} and \cite[Ch9-Thm.B1]{b_marshallOlkin}.
%, which follows from \cite[Ch9-Thm.B1]{b_marshallOlkin}
%Lemma3Kashyap_2003  \cite[Lem.3]{Kashyap_2003}

\bibliographystyle{ieeetr}%{ieeetr}
%    \AtEveryBibitem{%
%      \clearfield{da\mathbf{y}}%
%      \clearfield{month}%
%      \clearfield{endda\mathbf{y}}%
%      \clearfield{endmonth}%
%    }
%\kern-0.5em
\bibliography{\bibdirM/JNabrv,\bibdirM/bib_ayca,\bibdirMM/bib_randomMatrices,\bibdirM/bib_distributedEstimation,\bibdirM/bib_compressiveSampling2,\bibdirMM/bib_eh_M2M,\bibdirM/bib_energyHarvesting,\bibdirM/bib_csecuritySignal,\bibdirM/bib_robust,\bibdirMM/bib_robust_eh,\bibdirM/bib_optimization,\bibdirM/bib_books,\bibdirC/bib_ITrandomChannels,\bibdirMM/bib_mmse,\bibdirMM/bib_eh_intermittent,\bibdirMM/bib_eh_z,\bibdirMM/bib_mmse_comm,\bibdirMM/bib_fading,\bibdirMM/bib_hardware,\bibdirMM/bib_precoding_other,\bibdirMM/bib_eh_wsn,\bibdirM/bib_eh_prediction,\bibdirM/bib_linearEncoding}

%\bibdirM/bib_eh_ArrivalModels,

% %%%%%%%%%%%%%%%%%%%%%%%%%%%%%%%%%%%%
% \newpage
% ------
% \newpage
% \section{Other Material}
% \input{app_a_EH_intermittent_offlineJ}
% %%%%%%%%%%%%%%%%%%%%%%%%%%%%%%%%%%%%%%%%

\end{document}